\documentclass[fleqn,usenatbib]{mnras}
\usepackage[a4paper]{geometry}  

\usepackage{newtxtext,newtxmath}

\usepackage[T1]{fontenc}
\usepackage{ae,aecompl}
\usepackage{microtype}

\usepackage[utf8]{inputenc}
\usepackage{graphicx}	
\usepackage{amsmath}	
\usepackage{amssymb}	

\usepackage{multirow}
\usepackage{siunitx}    
\usepackage{booktabs}
\usepackage[dvipsnames]{xcolor}

\newcommand{\aref}[1]{\hyperref[#1]{Appendix~\ref{#1}}} 

\defcitealias{PythonTernary}{python-ternary}

\hypersetup{colorlinks=true,allcolors=teal}

\title[Velocity Dispersion and the Cosmic Web]{Large-Scale Velocity Dispersion and the Cosmic Web}

\author[M. Buehlmann \& O. Hahn]{
Michael Buehlmann$^{1}$\thanks{E-mail: michael.buehlmann@oca.eu} and 
Oliver Hahn$^{1}$
\\
$^{1}$Laboratoire Lagrange, Universit\'e C\^ote d'Azur, Observatoire de la C\^ote d'Azur, CNRS,\\
\phantom{$^{1}$} Blvd de l'Observatoire, CS 34229, F-06304 Nice cedex 4, France\\
}

\date{}

\pubyear{2018}

\begin{document}
\label{firstpage}
\pagerange{\pageref{firstpage}--\pageref{lastpage}}
\maketitle

\begin{abstract}
Gravitational collapse in cosmological context produces an intricate cosmic web of voids, walls, filaments and nodes. The anisotropic nature of collisionless collapse leads to the emergence of an anisotropic velocity dispersion, or stress, that absorbs most of the kinetic energy after shell-crossing. In this paper, we measure this large-scale velocity dispersion tensor $\sigma^2_{ij}$ in $N$-body simulations using the phase-space interpolation technique. We study the environmental dependence of the amplitude and anisotropy of the velocity dispersion tensor field, and measure its spatial correlation and alignment. The anisotropy of $\sigma^2_{ij}$ naturally encodes the collapse history and thus leads to a parameter-free identification of the  four dynamically distinct cosmic web components. We find this purely dynamical classification to be in good agreement with some of the existing classification methods. In particular, we demonstrate that $\sigma^2_{ij}$ is well aligned with the large-scale tidal field. We further investigate the influence of small-scale density fluctuations on the large-scale velocity dispersion, and find that the measured amplitude and alignments are dominated by the largest perturbations and thus remain largely unaffected. We anticipate that these results will give important new insight into the anisotropic nature of gravitational collapse on large scales, and the emergence of anisotropic stress in the cosmic web.
\end{abstract}



\color{black}  
\section{Introduction}
In the cold dark matter (CDM) paradigm of cosmological structure formation, all structure in the Universe forms through gravitational collapse from tiny metric perturbations seeded in an early inflationary phase of the Universe. Well after the dark matter particles become non-relativistic, their phase-space distribution function is well approximated by the cold limit, in which they occupy only a three-dimensional hypersurface in six-dimensional phase-space -- the Lagrangian submanifold. Its evolution is fully described by the Vlasov-Poisson system of equations which can be efficiently simulated using $N$-body methods, allowing us to reproduce the large-scale structure of the observed Universe to high accuracy \citep[e.g.][and many more]{Springel2006,Angulo2012,Potter2017}. 

At the earliest times, this submanifold coincides with three-dimensional space, but metric perturbations cause it to deform increasingly due to the growth of velocity perturbations reinforced non-linearly by self-gravity. At early times, every point in three-dimensional space is overlapped by exactly one point on the submanifold (the single-stream, or monokinetic regime). However, the growing velocity perturbations lead to shell-crossing at later times, resulting in multivalued velocities and the emergence of velocity dispersion and all higher moments of the Boltzmann hierarchy. Since dark matter in the standard CDM paradigm is collisionless, there is no rapid process that drives such multistreaming regions towards an isotropic Maxwell-Boltzmann distribution. The collapse of overdense regions proceeds at fixed scale through the well-known hierarchy of triaxial collapse: first to a planar structure, a \emph{pancake} or \emph{wall}, followed by collapse to a linear structure, a \emph{filament}, before the final axis collapses and a \emph{node} or \emph{halo} is formed \citep[a review of the various cosmic web environments can be found e.g. in][]{vandeweygaert2016}. This process is entirely hierarchical in the sense that in a universe with a full perturbation spectrum, one finds the intricate cosmic web structure \citep[e.g.][]{Bond1996} of haloes embedded in filaments, themselves embedded in walls on even larger scales.

To express this more formally, the Lagrangian submanifold can be parameterised by the Lagrangian coordinate $\mathbf{q}\in\mathbb{R}^3$ so that the phase-space distribution function can be written as
\begin{equation}
f_{\rm CDM}(\mathbf{x},\mathbf{v},t) =  \int \mathrm{d}^3 \mathbf{q} \, \delta_D\left(\mathbf{x}-\mathbf{x}(\mathbf{q},t)\right)\,\delta_D\left(\mathbf{v}-\mathbf{v}(\mathbf{q},t)\right),
\end{equation}
where $\mathbf{x}(\mathbf{q},t)$ and $\mathbf{v}(\mathbf{q},t)$ are the momentary position and velocity associated with $\mathbf{q}$ at time $t$. 

The time evolution of $\mathbf{x}(\mathbf{q},t)$ and $\mathbf{v}(\mathbf{q},t)$ at fixed $\mathbf{q}$ are simply the characteristics of the Vlasov equation conserving the phase-space density and following the canonical equations of motion \citep[cf.][]{Peebles1980}. The point-wise velocity distribution function (VDF) can then be thought of as evaluating $f_{\rm CDM}(\mathbf{x},\mathbf{v},t)$ at some fixed (Eulerian) point $\mathbf{x}$ in space. This evaluation implies counting how many $\mathbf{q}$s solve the implicit equation $\mathbf{x} = \mathbf{x}(\mathbf{q})$. The monokinetic regime is given by those regions where there is a single solution so that the VDF is a Dirac $\delta$-distribution (which is the zero-temperature limit of a Gaussian velocity distribution). In the multistreaming regime where there are multiple solutions, the VDF is a discrete sum over $\delta$-distributions. Hence, by Marcinkiewicz's theorem \citep{Marcinkiewicz1939}, there must be an infinite hierarchy of cumulants of the VDF. The monokinetic regime however is fully described by the continuity and the Euler equation. Only after shell-crossing, the second and all higher order cumulants emerge, so that an infinite hierarchy of fluid equations would have to be solved in the absence of collisions or other efficient relaxation processes suppressing the higher order cumulants. It is only on the smallest scales, i.e. inside haloes that are not dominated by recent accretion, that efficient (chaotic) mixing lets the velocity distribution approach a VDF relatively close to a Maxwell-Boltzmann distribution \citep[cf. e.g.][]{Hansen2006,Mao2013}.

In three dimensions, the emergence of the second and higher order cumulants happens only in those subspaces in which shell-crossing occurred due to the triaxial nature of the gravitational collapse. This anisotropic collapse should thus be reflected, at least on larger scales, in the second moment of the local VDF, which is the velocity dispersion tensor. We focus on this particular aspect in this paper by asking: \emph{(1) how does the anisotropic triaxial collapse of structure lead to anisotropic second moments of the VDF on large scales}, and, \emph{(2) to what degree is the anisotropic nature retained even in the presence of small-scale perturbations that will drive a gradual isotropisation in the deeply non-linear regime}. Studying the large-scale VDF is important for both our general understanding of cosmic structure formation and in particular for the implications on redshift-space distortions, since the three-dimensional positions of galaxies are always a sum of their positions and their line-of-sight velocities.

In this paper, we will measure the dark matter velocity dispersion based on the tessellation method introduced in \citet{Abel2012} and \citet{Shandarin2012}, which makes use of the fact that the distribution of CDM in the six-dimensional phase-space takes the form of a three-dimensional manifold that can be recovered well on large-scales in simulations. Using the particles of the $N$-body simulation as tracers of this distribution, the so-called dark matter sheet can be reconstructed by following the volumes spanned by neighbouring particles in Lagrangian space. This reconstruction method requires the knowledge of the primordial particle distribution, limiting its application mainly to cosmological $N$-body simulations. However, it has recently been shown that the dark matter phase space sheet can be reconstructed on large scales \citep[][]{Leclercq2017} from observational data, making the method also applicable to the local Universe.

The structure of this paper is as follows. First, in \autoref{sec:theory}, we discuss how gravitational collapse and the formation of the cosmic web lead to the emergence of velocity dispersion in the originally perfectly cold universe. Then, in \autoref{sec:simulations}, we provide a brief summary of the different dark matter simulations and the tessellation method that we use to measure the velocity dispersion tensor field. In \autoref{sec:results} we present our results on the amplitude and anisotropy measurements of the velocity dispersion field, including density dependences and two point correlations of the amplitude and alignments. We derive a natural and parameterless cosmic-web identification method from the anisotropy of the velocity dispersion tensor and study the density dependence and time evolution of the different cosmic web environments. Finally, we summarise our findings and conclude in \autoref{sec:conclusions}. We include a comparison with existing cosmic web identifiers in \aref{sec:cosmicweb_comparison}.


\section{The Emergence of velocity dispersion from shell-crossing}\label{sec:theory}
In this section, we discuss how velocity dispersion in the CDM fluid emerges as cosmological structures form -- on the largest scales in the form of the four well-defined distinct classes of cosmic web environments: voids, walls, filaments and nodes. We also explain how its strength and anisotropy are related to the scale and class of these structures.

  At early times (or on large-scales), the comoving dynamics of the CDM field can be described by the Zel'dovich approximation \citep[ZA,][]{Zeldovich1970}
  \begin{align}
    \bmath{x}(\bmath{q}, t) &= \bmath{q} + D_+(t) \nabla_q \phi(\bmath{q})
    &\bmath{u}(\bmath{q}, t) &= \dot{D}_+(t) \nabla_q \phi(\bmath{q}),
  \end{align}
  where $\bmath{q}$ and $\bmath{x}$ are the Lagrangian and the comoving Eulerian coordinates, $\bmath{u}$ are the velocities in comoving units, $\phi$ is a potential field proportional to the gravitational potential of the initial density perturbations and $D_+(t)$ is the linear theory growth factor.
  Since the phase-space coordinates $(\bmath{x}, \bmath{u})$ are a function of $\bmath{q}$, the dark matter fluid resides on a three-dimensional submanifold of the six-dimensional phase-space. 
  With $D_+(t) \to 0$ for $t \to 0$, the Lagrangian map $\bmath{q} \mapsto \bmath{x}$ is bijective at early times, and $\bmath{u}(\bmath{x})$ is single valued, hence there is no dark matter velocity dispersion in the early universe in this approximation.
  
  As time evolves, the dark matter fluid accelerates towards potential wells and the initial perturbations grow according to the growth function $D_+(t)$ until the dark matter sheet undergoes shell-crossing and enters the multistreaming regime. Since perturbations, and therefore the force fields, generally do not possess spatial symmetries, the collapse is anisotropic and happens at different rates along different axes. The principal axes of collapse correspond to the eigenvectors of the tidal field $T_{ij} = \partial^2 \phi/\partial q_i \partial q_j$.
  
  Once the perturbation is in the multistreaming regime, the velocity dispersion tensor\footnote{Note that in some literature, $\sigma_{ij}$ is also used to denote the velocity shear tensor $\Sigma_{ij} = \left(\partial\mathbf{v}_i / \partial\mathbf{x}_j + \partial\mathbf{v}_j / \partial\mathbf{x}_i\right) / 2$ \citep[e.g.][]{Peebles1980} and should not be confused with the velocity dispersion tensor defined here.} is defined as the variance of the velocities of the various streams at a given point, weighted by their respective local density on each stream,
  \begin{align} \label{eq:veldisp}
    \sigma_{ij}^2(\bmath{x}) &= \big< v_i(\bmath{x}) v_j(\bmath{x}) \big> - \big< v_i(\bmath{x}) \big>  \big< v_j(\bmath{x}) \big>,
  \end{align}
  where stream averaging is defined as
  \begin{equation}
\left<f(\bmath{x})\right> = \frac{\sum_k \rho^{(k)}(\bmath{x}) f^{(k)}(\bmath{x})}{\sum_k \rho^{(k)}(\bmath{x})}.
\end{equation}
Here, the index $k$ runs over the intersections of the dark matter sheet with position $\bmath{x}$, $\bmath{v} = a(t) \bmath{u}$ is the physical velocity, and $a(t)$ is the scale factor. In the following subsections, we will look at the evolution of the velocity dispersion as it emerges along the first axis of collapse and how the collapse along the subsequent axes are imprinted in $\sigma_{ij}^2$.
  
  \subsection{From one-dimensional collapse to the cosmic web}\label{sec:1dcollapse}
\begin{figure}
    \includegraphics[width=\columnwidth]{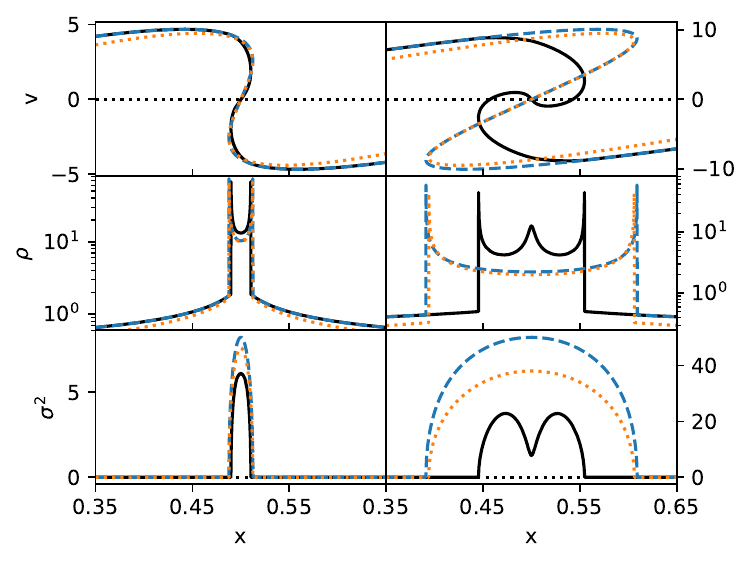}
    \caption{One-dimensional plane-wave collapse model of the dark matter sheet showing the phase-space distribution (top), the local density (middle) and the (comoving) velocity dispersion (bottom). The left panels show the system shortly after shell-crossing, whereas the system evolved further in the right panels. We compare the results from numerical integration (black, solid) with the ZA (blue, dashed) and the approximated plane wave perturbation (orange, dotted; see \aref{sec:1dtoy}). The ZA is correct up to shell-crossing and starts to deviate at later times, overestimating the size of the collapsed region and the velocity dispersion within.         
   The velocity dispersion in the fully evolved model peaks at $a \sim 1.4 a_\times$ and decays afterwards due to the growing density after subsequent further shell-crossings in the central region. As the ZA does not model this secondary collapse, the velocity dispersion increases until the expansion of the universe dominates the velocity field.}
    \label{fig:1dcollapse}
  \end{figure}

  To get an intuitive understanding of the velocity dispersion immediately after collapse, we will first look at a simplified model of a perturbation with a single mode $k$ in one dimension (a plane wave), $\phi(q) = A \cos(kq)$ with amplitude $A$. In the ZA, which in 1D is exact before shell-crossing, we can write 
  \begin{align}
    x(q, a) &= q - D_+(a) A k \sin(k q) & u(q, a) &= -\dot{D}_+(a) A k \sin(k q).
  \end{align}
  The mapping $q \leftrightarrow x$ is unique for $D_+(a) < D_+(a_\times)$, where shell-crossing occurs at $a_\times$. This time is defined by $D_+(a_\times) \equiv A^{-1} k^{-2}$, at which time the spatial derivative at $q=0$ vanishes and the wave collapses. \autoref{fig:1dcollapse} (top) illustrates the phase-space configuration of the plane wave shortly after collapse and at twice the collapse time, $a=2a_\times$. Whereas the ZA is still able to model the dark matter sheet in the first snapshot, it deviates strongly at later times and especially overestimates the width of the collapsed region.
  
  The Eulerian density shows the characteristic caustics at the border of the collapsed region where $\partial x/\partial q$ vanishes. In the ZA, these outer caustics are captured (albeit at the wrong locations) while the density peaks from subsequent shell-crossings at later times cannot be recovered. The velocity dispersion, zero in the monokinetic regime, rises strongly towards the inside of the collapsed region. Since the ZA does not capture subsequent shell-crossings due to secondary infall, it cannot correctly predict the long-term evolution of the velocity dispersion (cf. \autoref{eq:veldisp}). In the full non-linear solution, the central density increases together with the velocity dispersion, keeping the two in a dynamical equilibrium that cannot be described in the ZA.
  Nevertheless, the ZA provides a first estimate of the velocity dispersion immediately after shell-crossing. In \aref{sec:1dtoy} we derive an analytic solution for the lowest order polynomial expansion of a plane wave still leading to collapse (ZA+Taylor in \autoref{fig:1dcollapse}) and show that within the range of scales captured by our simulations, we expect the velocity dispersion measured shortly after collapse to increase with the scale of the perturbation.
  
  In three dynamical dimensions, collapse can proceed along a second and a third axis, first producing filaments, and finally haloes through shell-crossing from the multi-dimensional flow field (cf. \citet{Arnold1982, Arnold1982b, Hidding2014, Feldbrugge2018} for detailed discussions of the emergence of caustic singularities during anisotropic gravitational collapse and the nature of the multistreaming regions). The number of collapsed axes and their directions is directly imprinted in the velocity dispersion tensor, as velocity dispersion is nonzero only along the dimensions that have already collapsed. Studying the eigenvalues and eigenvectors of $\sigma_{ij}^2$ thus allows us to measure the advancement in structure formation and to identify the different components of the cosmic web. Since $\sigma_{ij}^2$ is a symmetric, positive semidefinite tensor, its associated eigenvalues $\lambda_i$ are real, and positive or zero. In theory, an eigenvalue of zero corresponds to an un-collapsed dimension along which no velocity dispersion has been generated. In this sense, a void has three zero eigenvalues, a pancake two, a filament one, and for a node all eigenvalues are different from zero. Naturally, due to finite numerical resolution and in a universe generated from random fluctuations on a multitude of scales, we will hardly find exactly vanishing eigenvalues. We discuss a natural way of separating the different regimes depending on the relative strengths of the eigenvalues in \autoref{sec:anisotropy}.


\section{Measuring large-scale velocity dispersion in $N$-body simulations} \label{sec:simulations}
To measure and analyse the properties of the dark matter velocity dispersion in the non-linear regime, we rely on numerical $N$-body simulations. In this section, we describe the initial conditions and parameters of the different cosmological simulations and the measurement of the velocity dispersion tensor from the $N$-body particle positions and velocities.

\subsection{Details of the simulations}
We have performed cosmological $N$-body simulations of CDM structure formation using the tree-PM code \textsc{gadget-2} \citep{Springel2005}, with initial conditions generated at redshift $z=99$ using \textsc{music} \citep{Hahn2011}. We use the \citet{Eisenstein1998a} transfer-function and cosmological parameters consistent with the Planck 2015 results \citep{Planck2015}, specifically $\Omega_m = 0.307$, $\Omega_b = 0.0486$ and $\Omega_\Lambda = 0.693$ for the density parameters, a Hubble parameter of $h = 0.6774$, normalisation $\sigma_8 = 0.816$ and spectral index $n_s = 0.9667$.
In addition, we carried out simulations for which we filter out small-scale structure in the initial condition, identical to warm dark matter (WDM) models. For these WDM initial conditions we truncate the transfer function according to \citet{Bode2001}
\begin{align}
  T_\text{WDM}(k) &= T_\text{CDM}(k) \left[ 1+(\alpha k)^2 \right]^{-5.0} \\
  \frac{\alpha}{h^{-1}\text{Mpc}} &= 0.05 \left(\frac{\Omega_m}{0.4}\right)^{0.15} \left(\frac{h}{0.65} \right)^{1.3} \left(\frac{m_\text{WDM}}{1\text{keV}} \right)^{-1.15},  
\end{align}
with a WDM particle ``mass'' of 250eV and 500eV, leading to truncation scales $\alpha=250 h^{-1} \text{kpc}$ and $\alpha=113 h^{-1} \text{kpc}$ respectively.
We define the characteristic scale where $T_\mathrm{WDM} / T_\mathrm{CDM} = 0.5$, the so-called \emph{half-mode scale} \citep[cf.][]{Schneider2012, Schneider2013}, which in our case are \smash{$k^\mathrm{hm}_\mathrm{WDM1} = 1.8$ $h$Mpc$^{-1}$} and \smash{$k^\mathrm{hm}_\mathrm{WDM2} = 0.85$ $h$Mpc$^{-1}$}, corresponding to tophat masses \smash{$M_\text{WDM}^\text{hm} = 2.2 \times 10^{12} h^{-1}\mathrm{M}_\odot$} and \smash{$M_\text{WDM2}^\text{hm} = 2.2 \times 10^{11} h^{-1}\mathrm{M}_\odot$}. The half-mode mass is where the WDM is expected to first affect the properties of haloes \citep{Schneider2012}. The scales that we are using are of course incompatible with observations, \citep[cf.][constraining the lower mass of sterile neutrinos in $\Lambda$WDM models obtained by Ly$_\alpha$ observations to $m_X \gtrsim 4.17$ keV corresponding to $M^\mathrm{hm} \sim 1.8 \times 10^8 h^{-1}\mathrm{M}_\odot$]{Yeche2017} but instead were tuned to correspond to roughly the non-linear scale $M_\ast$ at $z=0$, as well as to an approximately ten times smaller scale.
The details of the different simulations are summarised in \autoref{tab:sim}. We use the same amplitude for the power spectrum in the WDM initial conditions as the one derived from $\sigma_8$ in the CDM case, so that perturbations on large scales have identical amplitudes. Together with equivalent random seeds, all simulations have the same large-scale structure, up to some small back-reaction from small scales which are not present in the WDM runs.
The initial conditions are evolved to $z=0$, from where most of the following analysis is performed with the exception of the evolution of the collapse fraction for which 100 intermediate snapshots logarithmically distributed between $a=0.77$ and $a=1$ are used, and the power spectra which we also compute at $z=1$.

\begin{table}
  \begin{tabular}{@{}lccccc@{}}
    Simulation & $L_\text{box}$  & $N_p$ & $m_p$ &  $\epsilon$  & $m_\text{WDM}$\\
    & [$h^{-1}$Mpc] & & [$h^{-1}\mathrm{M}_\odot$] & [kpc] & [eV]\\
    \midrule
    300WDM1\_10 & 300 & $1024^3$ & $2.142 \times 10^9$ & 15   & 250\\
    300WDM1\_9 & 300 & $512^3$ & $1.714 \times 10^{10}$ & 30   & 250\\
    300WDM2\_10 & 300 & $1024^3$ & $2.142 \times 10^9$ & 15   & 500\\
    300WDM2\_9 & 300 & $512^3$ & $1.714 \times 10^{10}$ & 30   & 500\\
    300CDM\_10 & 300 & $1024^3$ & $2.142 \times 10^9$ & 6   & -\\
    300CDM\_9 & 300 & $512^3$ & $1.714 \times 10^{10}$ & 12   & -\\
  \end{tabular}
  
  \caption{Dark matter simulations used in this work and the corresponding parameters: box size $L_\mathrm{box}$, particle resolution $N_P$, particle mass $m_P$, force softening $\epsilon$ and WDM particle mass $m_\mathrm{WDM}$ used to truncate the initial power spectrum. All initial conditions were generated with the same random seed and the same normalisation of the power spectrum $A_s$, such that the large-scale structures are the same up to back-reaction from the small-scale structures in colder dark matter.}
  \label{tab:sim}
\end{table}


  \subsection{Measuring the velocity dispersion tensor in $N$-body simulations}
  To determine the velocity dispersion at any given point $\bmath{x}$, we reconstruct the fine-grained phase-space distribution of dark matter using the tessellation method described in \citet{Abel2012} and \citet{Shandarin2012}. We can think of particles in $N$-body simulations as tracers of the phase-space distribution. At early times before shell-crossing, they form a nearly uniform grid in configuration space. The unit cubes spanned by neighbouring particles can be decomposed into tetrahedra, which then form a natural tessellation of the phase-space dark matter sheet. This tessellation remains intact during the evolution of the system in the absence of collisions. Encoding the Lagrangian coordinate into the particle ids is enough to reconstruct the tessellation at any later time. There are different possibilities to decompose the unit cell into tetrahedra. For this paper, we use the equal volume, single coverage decomposition which was also used by \citet{Abel2012}.
    
  For a general discussion on interpolations on tetrahedra and phase-space projections of velocity fields, we kindly refer the reader to \citet{Hahn2014}. Here, we are mainly interested in the velocity dispersion tensor field which we can measure by computing the variance of the velocities of the tetrahedra intersecting a given point $\bmath{x}$ in configuration space:
  \autoref{eq:veldisp} becomes
  \begin{equation}
    \sigma^2_{ij}(\bmath{x}) = \frac{\sum_k \rho^{(k)} v_i^{(k)} v_j^{(k)}}{\sum_k \rho^{(k)}} - \frac{\sum_k \rho^{(k)} v_i^{(k)}}{\sum_k \rho^{(k)}}\frac{\sum_k \rho^{(k)} v_j^{(k)}}{\sum_k \rho^{(k)}}.
  \end{equation}
The fields $\rho^{(k)}$ and $v_i^{(k)}$ are interpolated linearly on the tetrahedra to the evaluation point using the values at the vertices. Alternatively, also the exact grid projection method of \cite{Powell2015} could have been used instead. The resulting field is in all cases volume weighted, which is very difficult to achieve by simple particle sampling since underdense regions are always poorly sampled \citep[as has already been pointed out by][]{Bernardeau1996}.

For our statistical result, we use the entire volume of the 300~$h^{-1}$Mpc boxes and compute the velocity dispersion on a $1024^3$ and $512^3$ grid.

\subsection{Density measurements and convergence}
 Whenever we use matter densities, we make estimates using both the cloud-in-cell (CIC) deposition algorithm \citep{Hockney1981} and the dark matter sheet tessellation \citep{Abel2012}. The reason for this is that the tessellation algorithm can overestimate densities when the distribution function cannot be well restored by linear interpolation between particles. This effect is known to be unproblematic when computing properties of velocity fields, since the density is used only as a relative weight for the different streams \citep{Hahn2014}. In fact, for velocity fields, the advantages of the phase space interpolation far outweigh the disadvantages, since particle based estimators are severely affected by shot noise in underdense regions, or artificially isotropised by kernels. We annotate the results with ``TESS'' or ``CIC'', depending on the method that was used. For all our measurements, we further compare results to those obtained with a lower resolution simulation. This allows an assessment of the degree of numerical reliability of our results.

\section{Results}\label{sec:results}

In this section, we present our measurements of the velocity dispersion tensor in the previously described simulations. We first look at the overall magnitude of the velocity dispersion, including environmental dependencies and spectral properties, before we explore its anisotropy and applications on the segmentation of the cosmic web. We then study the alignment of the velocity dispersion tensor with the local structure and the tidal field, and measure the influence of small-scale perturbations on the large-scale properties. 

\subsection{Visual impression}

\begin{figure*}
  \includegraphics[width=0.9\linewidth]{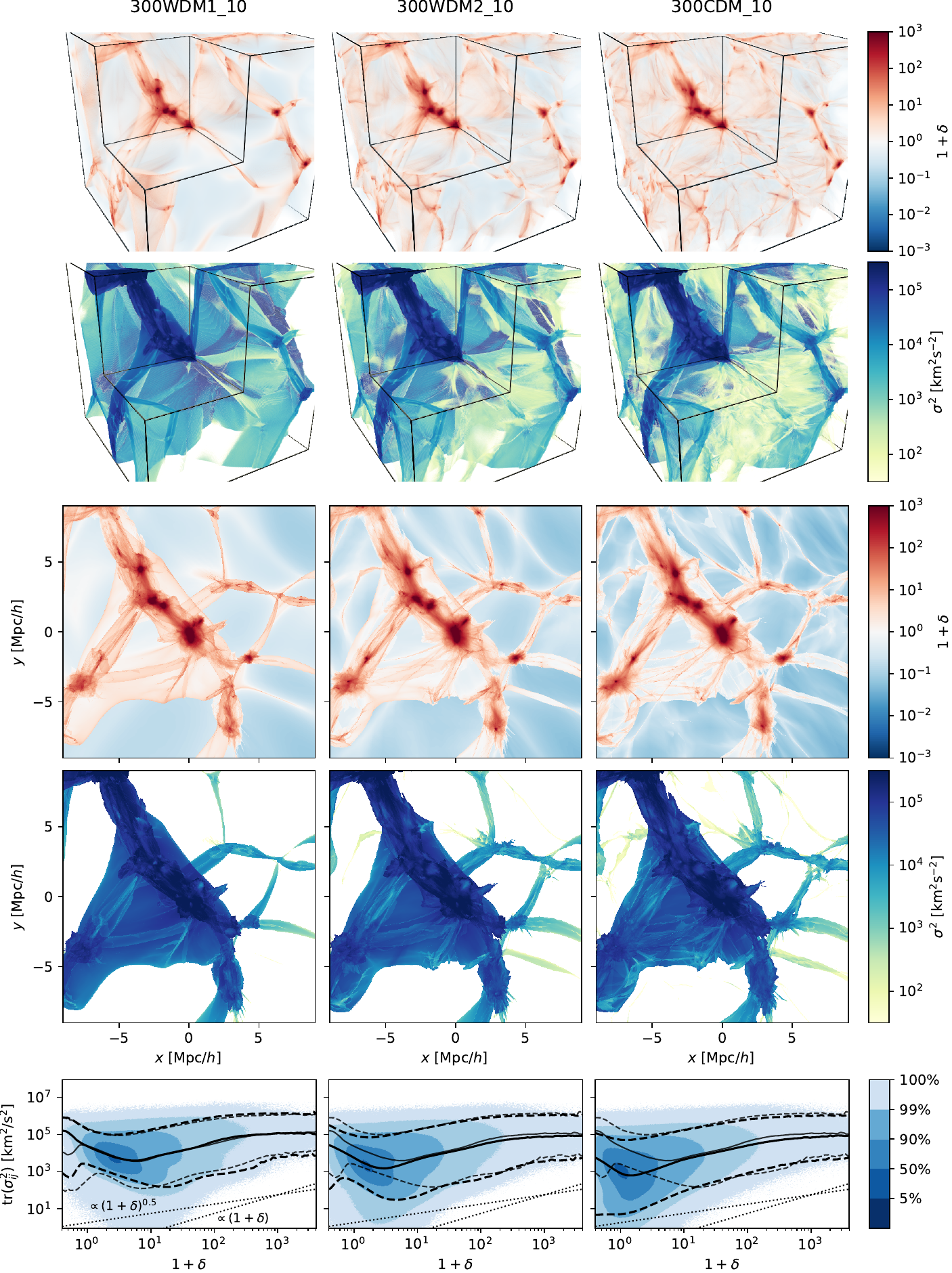}
  \caption{\textbf{Top rows:} Visualisation of the normalised density in the multistreaming regions (first and third row) and the amplitude of the velocity dispersion $tr(\sigma_{ij}^2)$ (second and fourth rows) for a 18~$h^{-1}$Mpc box around a $M_\mathrm{200c} = 8.9 \times 10^{13} h^{-1}\mathrm{M}_\odot$ halo. \textbf{Bottom row:} velocity dispersion amplitude $tr(\sigma_{ij}^2)$ -- density (TESS) distribution measured in the complete simulations, with the shaded areas indicating the 100\%, 99\%, 90\%, 50\% and the peak 5\% contours of the distribution. The black lines show the median (solid) and the 95\% interval for the distribution at a fixed density $1 + \delta$. Additionally we show the intervals obtained from the lower resolution simulations in lighter colours. The bottom dotted lines indicate a $(1+\delta)$ and $(1+\delta)^{0.5}$ slope as comparison.}
  \label{fig:visu1}
\end{figure*}

To give a visual impression of the cosmic velocity dispersion field and its spatial properties, we will first focus on the surroundings of a halo with mass $M_\textrm{200c} \sim 8.9 \times 10^{13} h^{-1}\mathrm{M}_\odot$. We compute the velocity and density fields from the tessellation of the dark matter sheet at points located on a uniform $512^3$ grid within a $18$~$h^{-1}$Mpc box centred at the halo. The halo we chose is massive enough to exist in all simulations. As can be seen from the density in the multistreaming regions shown in the top row of \autoref{fig:visu1}, the halo is embedded in the intersection of large walls with several other massive halos close-by. In the WDM simulations, low density walls and higher density filaments at the wall intersections are clearly visible. The highest density is reached in the central halo. Shifting the suppression scale for small-scale fluctuations towards lower masses extends the multistreaming web in previously uncollapsed regions and adds additional perturbations within the existing walls and filaments. Most notably, filaments begin to appear in walls, and nodes in filaments. We will define cleanly what we mean by \emph{walls}, \emph{filaments} and \emph{nodes} in \autoref{sec:anisotropy}.

\subsection{Magnitude of the large-scale velocity dispersion at $z=0$ and its density correlation} \label{sec:amplitude}

A natural measurement of the amplitude of the local velocity dispersion is the trace of $\sigma_{ij}^2$, which measures the sum of the dispersion along its main axes, $\mathrm{tr}(\sigma^2_{ij}) = \sum_i \lambda_i$ and essentially corresponds to the effective temperature of the dark matter due to gravitational collapse. The second and fourth rows of \autoref{fig:visu1} illustrate the velocity dispersion amplitude in the three simulations. Starting from the 300WDM1 simulation, we can clearly see walls separating large volumes with no velocity dispersion. These regions, the cosmic voids, have not collapsed and are thus still in the monokinetic single-stream regime. Among the wall regions, the velocity dispersion increases with the size and thickness of the structures. This is consistent with our one-dimensional collapse model presented in \aref{sec:1dtoy}, predicting that larger-scale perturbations also have higher velocity dispersion after collapse and result in wider collapsed regions. In the centre of the cross sections of the large walls, we can see additional finer structures with lower velocity dispersion. These structures originate from the secondary collapse in a direction perpendicular to the wall, causing enhanced densities and suppressed velocity dispersion due to the higher weight of the inner streams (cf. \autoref{fig:1dcollapse} at $a=2.0a_\times$).

The major structures remain remarkably similar in all simulations with only little change in the strength of the velocity dispersion. Decreasing and removing the suppression of small-scale perturbations in the initial conditions naturally adds multistreaming structures in the voids \citep[cf. e.g.][for a detailed discussion of this aspect]{Stuecker2018}. These fine structures are small in size and width and have relatively low velocity dispersion. The additional small-scale structures within existing collapsed regions have a relatively small effect on the measured velocity dispersion as it is dominated by the large-scale modes. We will further discuss and quantify the influence of small perturbations in \autoref{sec:alignment}.

To further investigate the relationship between density and velocity dispersion, we compute $\mathrm{tr}(\sigma^2_{ij})$ on the full box and plot its distribution with respect to the density (TESS) measured in the multistreaming regions (remember that the velocity dispersion vanishes exactly in single stream regions). The results are plotted in the bottom row of \autoref{fig:visu1}. The shaded regions show the 100\%, 99\%, 90\%, 50\% and the peak 5\% contours and the black lines the median and the 95\% interval of the distribution at a specific density $1 + \delta$. Additionally, we include the results from the lower resolution simulation to test for convergence. We find consistent results at high densities but deviations especially for the CDM simulation in low density environments, where the lower resolution simulation fails to capture the collapsed small-scale fluctuations with low velocity dispersion and low densities.

We find that above a density of $\delta \sim 4$ in the 300WDM1 realisation and $\delta \sim 1$ in the 300CDM\_10 simulation the velocity dispersion is positively correlated with density, with roughly $\mathrm{tr}(\sigma^2_{ij}) \propto (1+\delta)^{\alpha}$ and $\alpha \sim 0.5-1$. The correlation is stronger in the 300CDM simulation due to the additional low density, low velocity dispersion regions which do not exist when small-scale fluctuations are suppressed. At low densities this trend is reversed and the velocity dispersion increases towards the few collapsed regions below mean density. This is most likely due to the collapsed regions with the lowest densities -- walls or pancakes -- but which still have high velocity dispersion if they originate from a large-scale mode as can be seen from the visualisations.

The volume distributions of the multistreaming regions peak at $\delta \sim 1-3$ and $\mathrm{tr}(\sigma^2_{ij}) \sim 10^{3} - 10^{4}$, depending on the truncation scale of the small-scale structure (higher densities and higher velocity dispersion in the truncated simulations). The presence of small-scale fluctuations in the 300WDM2 and 300CDM simulations does not affect the distribution at the high end of the velocity dispersion distribution, but adds structures with low velocity dispersion and low to medium ($\delta \sim 10$) density. This is consistent with the observation of added small-scale structure with low velocity dispersion in previously uncollapsed regions but persistently high velocity dispersion within the large-scale structures.


\subsection{Two-point statistics of the cosmic velocity dispersion} \label{sec:power}
As one can already see from the 3d-visualisations in \autoref{fig:visu1} and \autoref{fig:visu2}, the velocity dispersion is spatially correlated, both in amplitude and in direction. First, we will focus on amplitude auto- and cross-correlations and will focus on directional correlations below in \autoref{sec:anisotropy}. Since the velocity dispersion vanishes in single-stream regions, the field is not defined everywhere in space so that the resulting two-point statistics will also include a strong signal of the size and shape of multistreaming regions.

To analyse the spatial clustering of the velocity dispersion, it is useful to measure its autocorrelation and cross-correlation with the density field in Fourier space. The density and velocity dispersion power spectra and the corresponding cross-spectrum are given by
\begin{align}
  \left<\delta(\bmath{k})\delta(\bmath{k}')^\ast\right> &= P_{\delta\delta}(k) \; \delta_D(\bmath{k}-\bmath{k}') \\
  \left<\sigma^2(\bmath{k})\sigma^2(\bmath{k}')^\ast\right> &= P_{\sigma^2\sigma^2}(k) \; \delta_D(\bmath{k}-\bmath{k}') \\
  \left<\delta(\bmath{k})\sigma^2(\bmath{k}')^\ast\right> &= P_{\delta \sigma^2}(k) \; \delta_D(\bmath{k}-\bmath{k}'),
\end{align}
with $\sigma^2=\mathrm{tr}(\sigma_{ij}^2)$. We deconvolve the density field (CIC) with the CIC assignment kernel $W_\mathrm{CIC}(\bmath{k}) = \prod_i \mathrm{sinc}^2(k_i / 2 k_\mathrm{Ny})$ to correct for the smoothing effect of the mass assignment scheme close to the Nyquist wave number $k_\mathrm{Ny}=N \upi / L$, and de-alias the measured density power spectrum by interlacing the original field with a grid shifted by half a cell size in all directions \citep[cf. ][]{Sefusatti2016}.

\begin{figure*}
  \includegraphics[width=\linewidth]{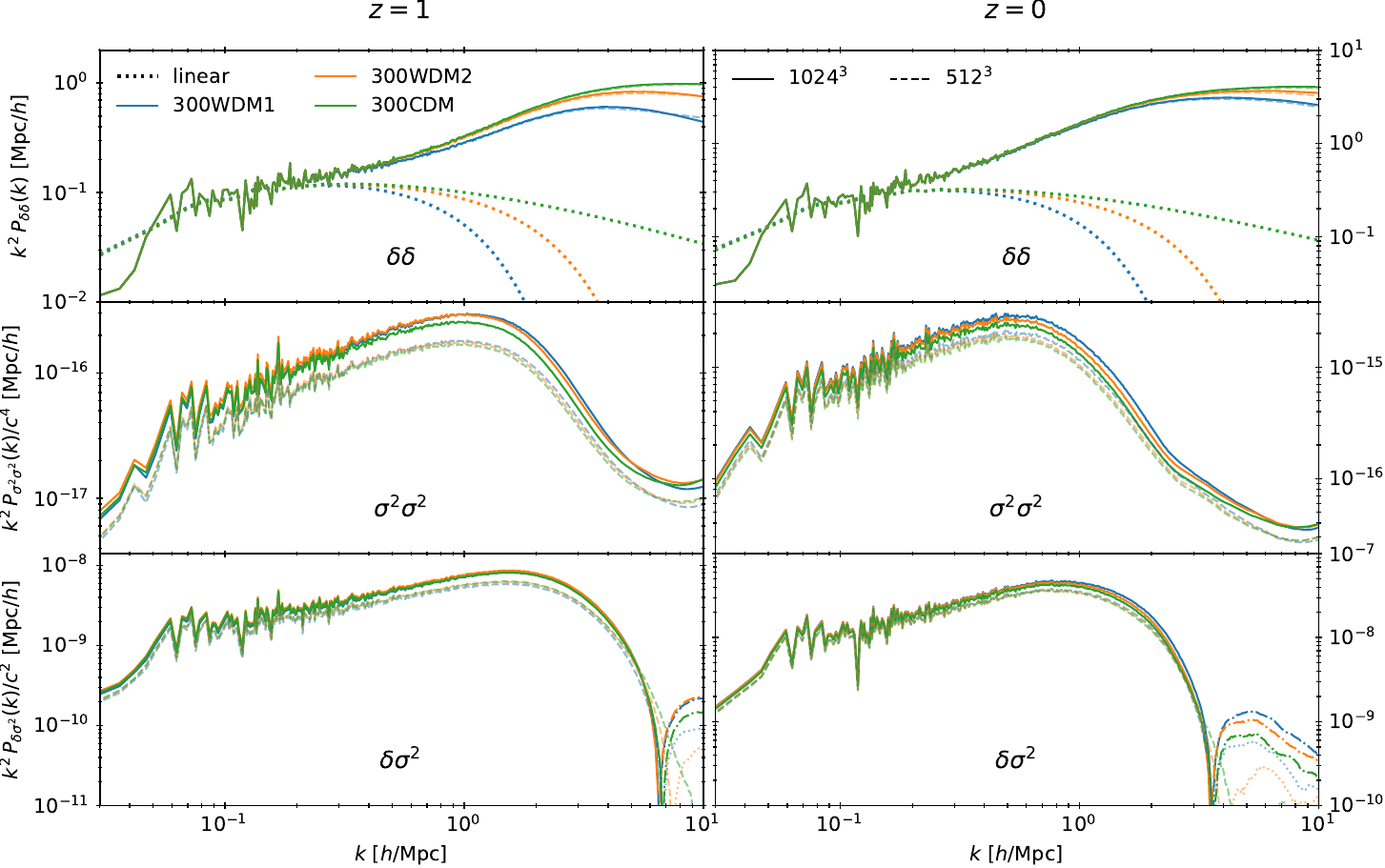}
  \caption{Power spectra of the density (top), velocity dispersion $\sigma^2 = \mathrm{tr}(\sigma_{ij}^2)$ (middle) and cross correlation (bottom) at redshift 1 and 0.  Dashed lines show the lower resolution realizations and negative values in the case of the cross-spectrum at large $k$ are indicated with dash-dotted and dotted lines.}
  \label{fig:vd_power}
\end{figure*}

\autoref{fig:vd_power} shows the measured power spectra at redshift $z=1$ and $z=0$ for the three simulations. Starting from the top, we notice that at both redshifts the matter power spectrum is enhanced at small scales compared to the linear power spectrum due to the non-linear growth of structures. As expected from the set-up of the simulations, the amount of small-scale clustering is dependent on the truncation scale of the initial power spectrum. A comparison between the low and high resolution simulations shows that the results are well converged.

For the velocity dispersion power spectrum the situation is somewhat different: even at the largest scales, we find a measurable offset of $\sim20$ per cent in amplitude between the 300WDM1, 300WDM2 and 300CDM simulations. A comparison with the lower resolution simulations shows that none of the measurements are perfectly converged. This lack of fast convergence is very reminiscent of the convergence properties of the vorticity power spectrum, where various studies have found that the non-linear scale has to be very well resolved \citep{Pueblas2008,Hahn2014}. Just as the vorticity, $\sigma_{ij}^2$ is only non-zero in multi-stream regions and thus more strongly affected by resolution than those quantities that are non-zero also in the monokinetic regime and thus defined everywhere in space.

Since shell-crossing occurs predominantly in overdense regions, the velocity dispersion is highly correlated with the density field on large scales. On the largest scales, we find that $P_{\sigma^2\sigma^2}(k) \propto k^{-1}$, with a sharp drop on small scales. The same holds for the cross-spectrum, which however becomes negative above $k \sim 6$~$h$Mpc$^{-1}$ at $z=1$. The inversion moves towards larger scales at later times, with $k \sim 3.5$~$h$Mpc$^{-1}$ at $z=0$. This is a signature of the largest collapsed structures and has also been observed in the density -- velocity divergence cross-spectrum \citep{Hahn2014, Jelic-Cizmek2018} and the velocity divergence -- velocity dispersion cross-spectrum \citep{Jelic-Cizmek2018} at similar scales. The transition from correlation to anticorrelation at small scales is of course consistent also with the two-dimensional density-dispersion histograms we presented in the previous section. We note that the scale of anticorrelation between overdensity and velocity dispersion is independent of the type of simulation and nearly independent of the resolution, indicating further that it originates from the largest collapsed structures. This phenomenon thus plausibly originates from the outer shells of the largest collapsed structures and we therefore expect it to be intimately related to the splash-back radius of galaxy clusters \citep[cf.][]{Gill2005,More2015,Mansfield2017}, which denotes the outer caustic in isotropically collapsed systems (what we will call `nodes' below).


\subsection{The anisotropy of the velocity dispersion tensor -- from voids to walls to filaments to nodes} \label{sec:anisotropy}
\begin{figure*}
  \includegraphics[width=\linewidth]{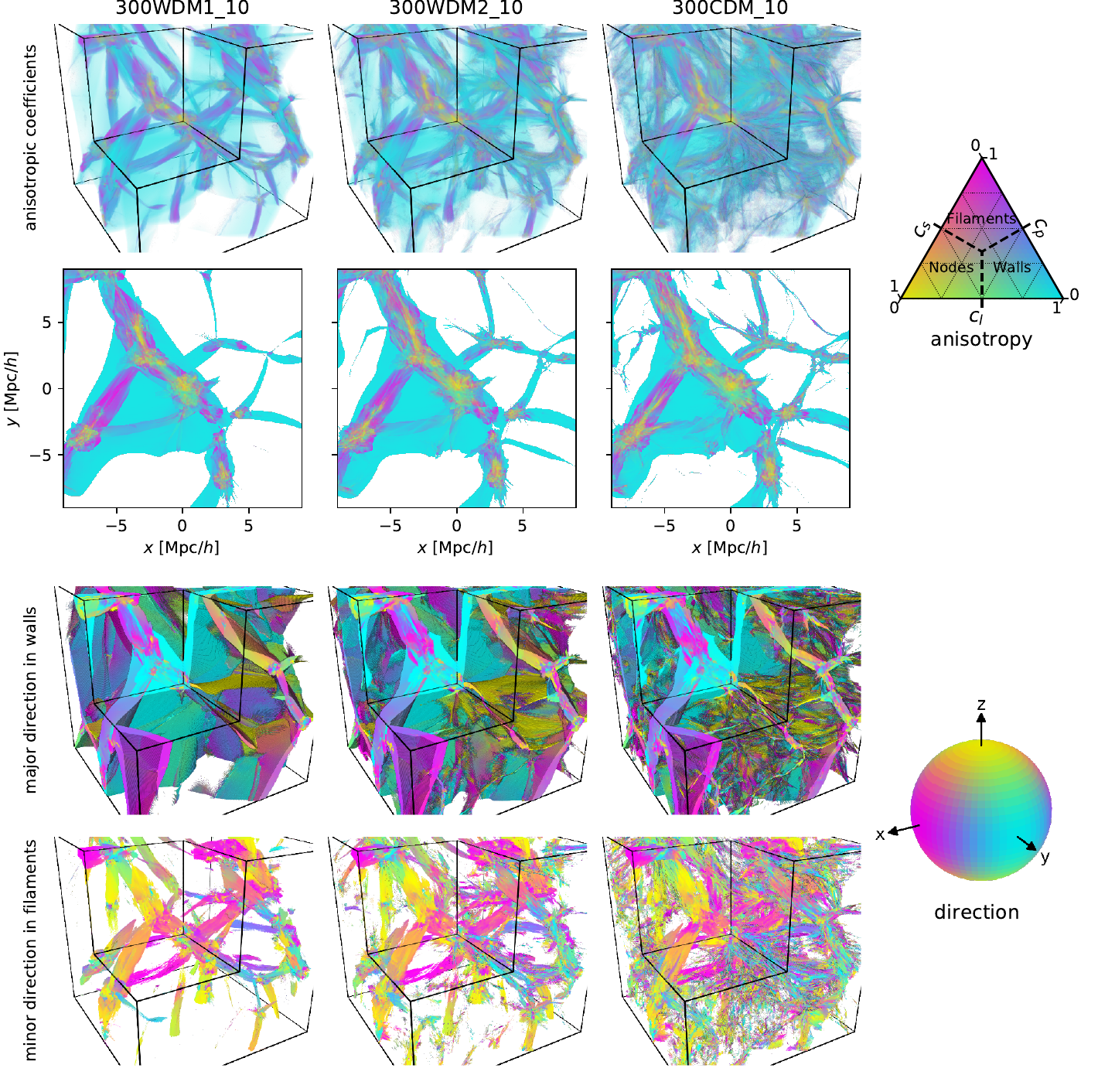}
  \caption{\textbf{Top:} the anisotropy coefficients $c_l$, $c_p$ and $c_s$ (top rows) for a $\sim 18$~$h^{-1}$Mpc box around a halo for the 300WDM1\_10 (left), 300WDM2\_10 (middle) and 300CDM\_10 (right) simulation. The anisotropy is colour-coded according to the ternary diagram, with linear, planar and symmetric dominant regions in cyan, magenta and yellow respectively. \textbf{Bottom:} Direction of the major eigenvector of $\sigma_{ij}^2$ within wall-like regions and the minor eigenvector within filament-like regions. The characteristic directions are colour-coded according to the unit sphere shown on the right and they are perpendicularly aligned within wall-like regions and parallel to the extent of filaments.}
  \label{fig:visu2}
\end{figure*}

In \autoref{sec:theory} we argued that the progression of anisotropic collapse from walls to filaments to nodes should be reflected directly in the tensor $\sigma_{ij}^2$, absent any strong isotropisation processes. In an idealized setting we would expect that uncollapsed and collapsed axes should correspond to vanishing and non-zero eigenvalues of the velocity dispersion tensor, respectively. Naturally, an exact vanishing of eigenvalues is unlikely in a numerical setting, and due to overlapping perturbations on various scales. We therefore derive three dimensionless quantities from the eigenvalues $\lambda_1 > \lambda_2 > \lambda_3 \ge 0$ of $\sigma_{ij}^2$ which capture the relative strengths of the collapsed dimensions (reflecting thus dominantly one-, two- or three-dimensional collapse):
\begin{enumerate}
\item the linear anisotropy $c_l = (\lambda_1 - \lambda_2)/(\sum \lambda_i)$,  
\item the planar anisotropy $c_p = 2(\lambda_2-\lambda_3)/(\sum \lambda_i)$, and 
\item the spherical anisotropy (or isotropy) $c_s = 3\lambda_3/(\sum \lambda_i)$. 
\end{enumerate}
Note that by construction $c_l + c_p + c_s = 1$, and hence these quantities parametrize a so-called barycentric space with three extrema ($c_s = 1$: fully symmetric, $c_p = 1$: symmetric along two axes and zero along the third, $c_l = 1$: dispersion only along one axis) and can be represented in a ternary diagram, as shown on the right of \autoref{fig:visu2}. We can divide this diagram into three parts depending on the dominant parameter and decide if a region is either
\begin{enumerate}
\item linear-anisotropic $\Leftrightarrow\,\, c_l$ is dominant $\,\Leftrightarrow\,\,$ `wall'-like, 
\item planar-anisotropic $\Leftrightarrow\,\,c_p$ is dominant $\,\Leftrightarrow\,\,$ `filament'-like,
\item isotropic $\Leftrightarrow\,\, c_s$ is dominant $\,\Leftrightarrow\,\,$ `halo'-like. 
\end{enumerate}
For a more sophisticated classification of the cosmic web, one might consider different segmentations of the anisotropy triangle, for example, by classifying filaments as regions which have a planar anisotropy larger than a small threshold, above which one assumes that collapse along the second axis has started. An analogous argument can be made for the isotropic component and nodes. We leave the investigation of these advanced classifications for a later study and define walls, filaments and nodes depending on the dominant anisotropic parameter which avoids the introduction of additional parameters. Note that the multiscale nature of the cosmic web means that filaments can be embedded in walls and nodes in filaments. Since this method unqiquely identifies the environment at a specific point in space by its dominant anisotropic parameter, it does not resolve this hierarchical structure and the classification depends on the amount of small-scale perturbations. To identify the cosmic web on various scales, additional smoothing steps (either by supressing small-scale fluctuations in the initial conditions or by post-processing) are required.

\begin{figure}
  \includegraphics[width=\columnwidth]{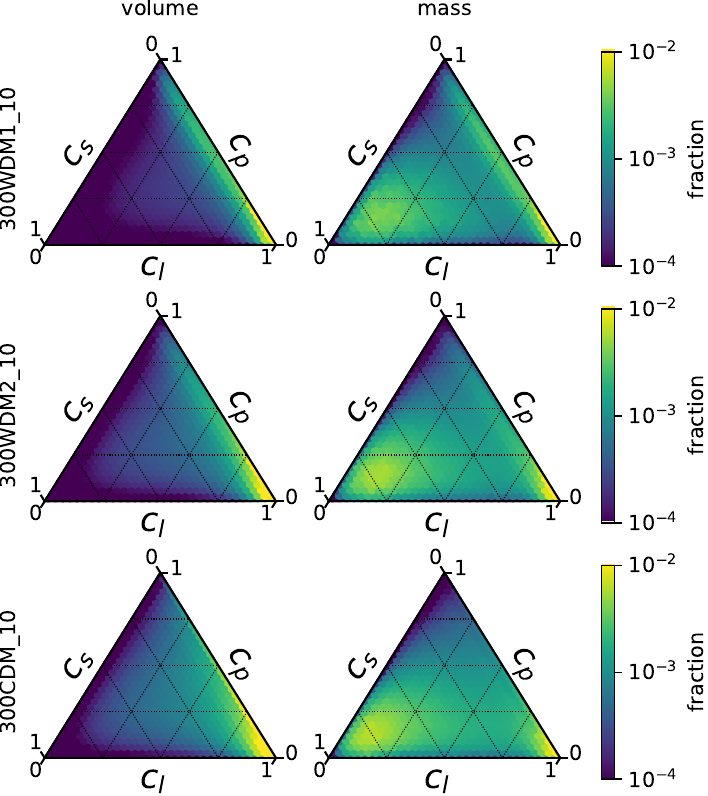}
  \caption{Distribution of volume (left) and mass (right) in multistreaming regions according to the anisotropic coefficients $c_l$, $c_p$ and $c_s$. In most of the volume, the velocity dispersion is linear-anisotropic, with some fraction being between linear- and planar-anisotropic. The mass weighted distribution is bimodal with a second peak at high isotropy due to the contribution of the high density regions.}
  \label{fig:ani_dist}
\end{figure}

We compute the anisotropy parameters for each volume cell in the multistreaming region and show the resulting volume and mass distribution in \autoref{fig:ani_dist} for the 300WDM1\_10, 300WDM2\_10 and 300CDM\_10 simulations. The largest part of the volume in all simulations has a highly  linear-anisotropic velocity dispersion and hence is in wall-like structures. This is  more pronounced in the WDM1 case, where 83\% of the multistreaming volume has a planar anisotropy coefficient smaller than 0.25 and only a small fraction ($\sim 3$\%) has an isotropic coefficient larger than 0.25. In the CDM simulation a larger fraction of the collapsed volume is isotropised, with 26\% of the volume having $c_p > 0.25$ and 10\% having $c_s > 0.25$. Interestingly, we can see that the large volume fraction with vanishing spherical anisotropy in the 300WDM1 realization moves away from the $c_s=0$ line and becomes more isotropised by the small-scale perturbations in the 300CDM simulation.   

\begin{figure*}
  \includegraphics[width=\linewidth]{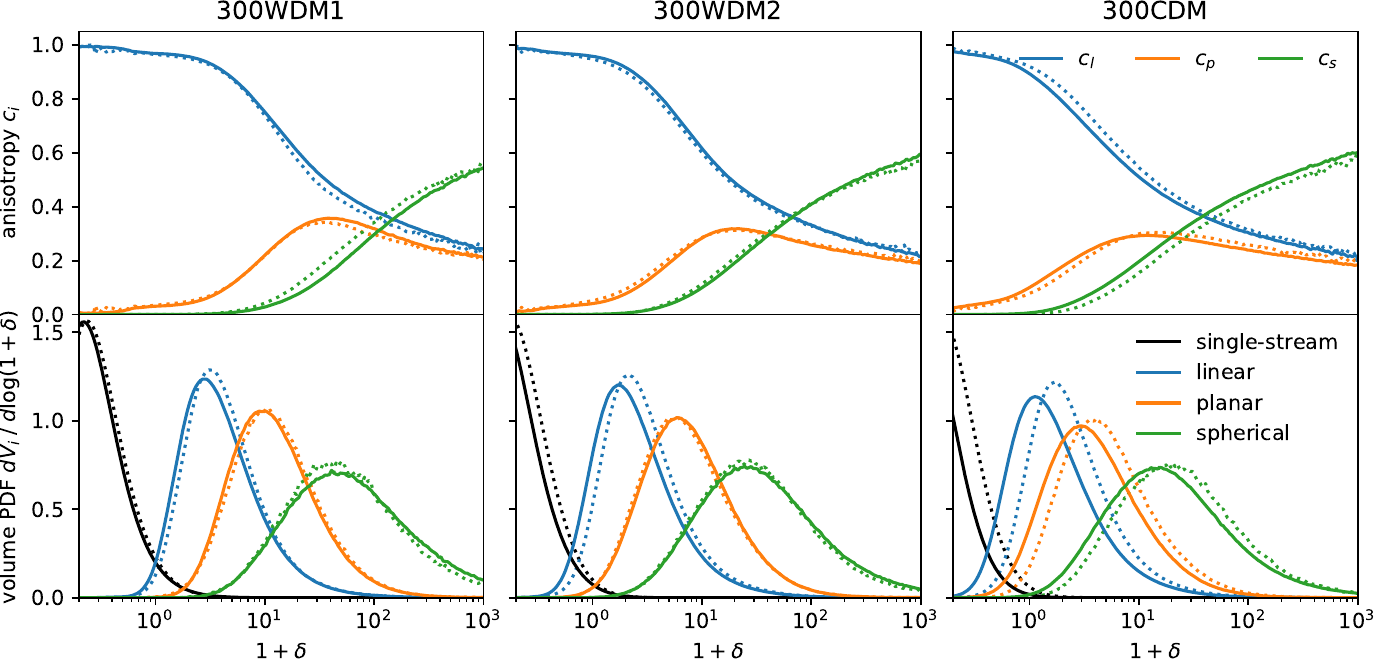}
  \caption{\textbf{Top:} Average value of the anisotropy coefficients depending on the local density (TESS) for the 300WDM1 (left), 300WDM2 (middle) and  300CDM (right),  simulation at high (solid) and low (dotted) resolutions.  \textbf{Bottom:} Relative distribution of regions with predominately linear anisotropy (walls), planar anisotropy (filaments) and spherical anisotropy (nodes) depending on the local density. We also include the single-stream distribution that falls within the shown density range.}
  \label{fig:density_aniclass}
\end{figure*}

Looking at the mass-weighted distribution, we find a second peak at large $c_s$ originating from high density regions which are predominantly located in fully collapsed structures (cf. \autoref{fig:density_aniclass}). We find 17\% (26\%) of the collapsed mass being in regions with a planar coefficient larger than 0.25 and 42\% (60\%) in regions with an isotropic coefficient larger than 0.25. The results for the 300WDM2\_10 simulation are located between the the other two simulations.

\begin{table}
  {\scriptsize
  \begin{tabular}{@{}rlcccc@{}}
    &Simulation & uncollapsed & walls & filaments &  nodes \\
    & & \% & \% & \% & \%\\
    \midrule
    \multirow{3}{*}{\rotatebox[origin=c]{90}{volume}}
    &  300WDM1 & 95 (95) &  4.9 ( 4.5) &  0.5 ( 0.4) &  0.1 ( 0.1) \\
    &  300WDM2 & 92 (93) &  7.4 ( 6.3) &  0.7 ( 0.7) &  0.2 ( 0.2) \\
    &  300CDM  & 89 (91) &  9.6 ( 7.4) &  1.2 ( 1.0) &  0.4 ( 0.3) \\
    \midrule
    \multirow{3}{*}{\rotatebox[origin=c]{90}{mass}}
    &  300WDM1 & 41 (48) & 33 (31) & 11 ( 9) & 16 (13) \\
    &  300WDM2 & \emph{30 (40)} & 35 (33) & 10 (10) & \emph{24 (17)} \\
    &  300CDM  & \emph{24 (37)} & 36 (34) & 11 (10) & \emph{30 (19)} \\
    \\
    & & $\bar{\delta}$ & $\bar{\delta}$ & $\bar{\delta}$ & $\bar{\delta}$ \\
    \midrule
    \multirow{3}{*}{\rotatebox[origin=c]{90}{density}}
    & 300WDM1 & -0.6 (-0.5) &  5.7 ( 5.9) &   22 (  21) &  \emph{204 ( 125)} \\
    & 300WDM2 & -0.7 (-0.6) &  3.8 ( 4.3) &   14 (  13) &  \emph{115 (  78)} \\
    & 300CDM  & -0.7 (-0.6) &  2.7 ( 3.6) &    8 (   9) &   69 (  63) \\    
  \end{tabular}
  \caption{Top: mass and volume fractions in the single- and multistreaming regions. The multistreaming regions are split by the dominant anisotropy parameter into \emph{linear} (wall-like), \emph{planar} (filament-like) and \emph{isotropic} (node-like) environments. The percentages are computed from the $1024^3$ particle realisations using the tessellation density estimate, with the CIC densities showing consistent results. 
  Bottom: mean density of the individual environments.
  Values from the $512^3$ simulations are given in parentheses for comparison and highlighted in italic if they show a strong discrepancy.}
  \label{tab:volmass_dist}
  }
\end{table}

In \autoref{tab:volmass_dist} we list the volume and mass fractions of the $c_p$, $c_l$, $c_s$ dominant and of the single-stream regions for the different simulations, including the lower resolution simulations to check for convergence. We note that convergence of results in the CDM limit is generally a non-trivial question since in the perfectly cold limit, virtually all structure on the investigated scales should be in haloes \citep[cf.][]{Stuecker2018}. This can be seen from the CDM volume and mass fraction in node-like regions that is larger in the higher resolution simulation.

The fraction of volume and mass in multistreaming regions increases with colder simulations, consistent with the additional small-scale structures observed in \autoref{fig:visu1}. The uncollapsed single-stream regions remain the dominant fraction of the volume, whereas most mass can be found in collapsed regions. Each of the three multistreaming regions gains volume by adding small-scale structures, but only the mass fraction in node-like regions increases significantly, while walls filaments remain roughly constant.

Previous studies on volume and mass fractions in the cosmic web have found a wide range of values \citep[a recent comparison can be found in][also see the comparison in \autoref{tab:volmass_comparison}]{Libeskind2018}. These sometimes large discrepancies are the result of different classification criteria and complicate the comparison of our results. Closely related to our method is i) the phase-space tessellation by \citet{Shandarin2012} (also see \citealt{Ramachandra2015}) which uses stream number thresholds to identify the different components, i.e. voids: $n_\mathrm{stream}=1$, walls: $3 \leq n_\mathrm{stream} < 17$, filaments: $17 \leq n_\mathrm{stream} < 90$ and haloes: $n_\mathrm{stream} \geq 90$, and ii) the phase-space folding detection method \textsc{origami} \citep{Falck2012}, finding the individual axes along which the Lagrangian particle ordering is inverted. Since we use the same single-stream definition as i), our mass and volume fractions for cosmic voids are comparable to their results. We note that the reported void volume fractions using this technique are consistently $\sim90$\% but the mass fractions vary strongly (23\% by \citealt{Shandarin2012}, 32\% by \citealt{Ramachandra2015} and 56\% by \citealt{Libeskind2018}). The void volume fraction reported in \citet{Libeskind2018} for \textsc{origami} is lower (70\%), with a larger volume fraction classified as nodes (7.4\%) and filaments (6.4\%). The mass is equally more attributed to nodes (50\%), but less to walls (14\%).

Compared to further methods, we find our volume and especially mass fraction of filament-like regions to be towards the lower end of the wide range of reported values. Results for the mass fraction using stream number thresholds range from 10\%-20\%, with other methods assigning up to 50\% \citep[\textsc{nexus+},][]{Cautun2014} and 60\% \citep[\textsc{DisPerSE},][]{Sousbie2011, Libeskind2018} of the total mass to filaments . On the other hand, the total mass fraction of wall-like regions is at the upper level of previous studies (13\%-33\%), with our results being comparable to \citet{Shandarin2012}.

To further investigate the density dependence of the anisotropic parameters, we measure their mean value as a function of the local density (TESS) $\delta(\bmath{x}) + 1$. The results are shown in the top panels of \autoref{fig:density_aniclass} and a comparison with the low resolution simulation (dotted) shows that they are fairly well converged in the simulations with truncated small-scale perturbations. In collapsed regions close to the mean density, the linear anisotropy parameter is strongest, whereas high-density regions have a predominantly isotropic velocity distribution. The planar anisotropic coefficient peaks around $\delta \sim 10$ in the CDM case and $\delta \sim 30$ in the WDM1 simulation and decreases for both smaller and higher overdensities. The simulations differ mainly in the low density regime, which is almost purely linearly anisotropic in the 300WDM1 simulation, but has a small planar anisotropic contribution in the 300CDM simulation. Overall, small-scale perturbations lower the density threshold at which shell-crossing along the second and third axes can occur and thus at which the planar anisotropy and isotropy become measurable. In the 300CDM simulation, this threshold remains unresolved due to initial density fluctuations at arbitrarily small scales.

Labelling each cell by its dominant anisotropic parameter, we can examine the density distribution of linear-, planar- and spherical-anisotropic multistreaming regions (walls, filaments and nodes). The results are shown in the lower panel of \autoref{fig:density_aniclass} and the mean density of each environment can be found in \autoref{tab:volmass_dist}. In agreement with the measured mean anisotropic coefficients, wall-like regions are predominantly in low density environments, followed by filament-like and node-like regions (this hierarchy also directly follows from the study of the initial shear tensor by calculating the probability of the eigenvalue signatures depending on the local overdensity, cf. \citet{Pogosyan1998}). In the 300WDM\_10 simulation, the distributions peak at $\delta \sim 3$, $\delta \sim 10$ and $\delta \sim 50$, respectively. In the case of the colder simulation, the distributions shift towards lower densities due to the additional small-scale perturbations ($\delta \sim 1$, $\delta \sim 3$ and $\delta \sim 20$, respectively, for the 300CDM\_10 simulation). A comparison to the low resolution simulation (dotted) shows qualitatively consistent results for both WDM simulations but a shift towards higher densities in the 300CDM\_9 simulation due to the unresolved small-scale structure. The density distributions reported by other cosmic web finders vary strongly (cf. \aref{sec:cosmicweb_comparison} and also the extensive comparison of cosmic web finders by \citet{Libeskind2018}), which makes a direct comparison difficult since most structure finders define the various environments in fundamentally different ways.


\subsection{Evolution of mass and volume fractions of structures over cosmic time} \label{sec:evolution}

Most of our results above have been obtained at $z=0$. In order to complement this momentary picture at late time, we investigate in this subsection the evolution of the collapsed regions of the respective morphologies over cosmic time.

\begin{figure}
  \includegraphics[width=\columnwidth]{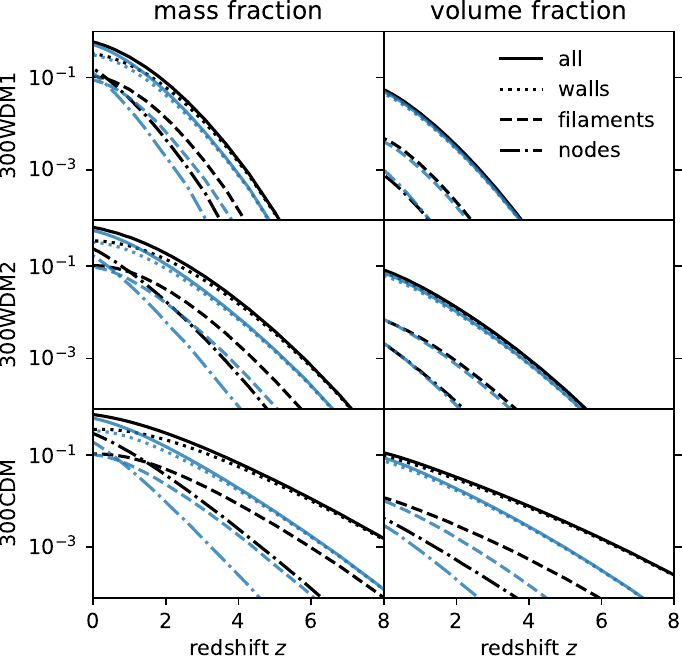}
  \caption{Evolution of the fraction of the mass (left) and volume (right) that has collapsed (solid lines) and the subdivision according to the anisotropic shape of the velocity dispersion tensor (dotted and dashed lines) for the CDM and WDM simulations (rows). To check for convergence, we include the results for the low resolution simulations (blue). The cell densities have been obtained by the CIC algorithm.}  
  \label{fig:collapse_history}
\end{figure}

From the theory of anisotropic collapse, we expect the first multistreaming regions to have linear-anisotropic velocity dispersion. Planar-anisotropic and isotropic velocity dispersion emerge at a later stage when the wall-like structures collapse along the second and third axes. As the collapse time depends on the amplitude and scale of the perturbation (cf. \aref{sec:1dtoy}), more single-streaming volume will continuously enter the linear-anisotropic regime and collapse further. In \autoref{fig:collapse_history}, we measure the volume and mass fraction that has collapsed at different times during our simulation. As the perturbations with the highest overdensities enter the multistreaming regime first, a significant fraction of the total mass can be found in collapsed regions before the same fraction of volume has collapsed. The multistreaming mass fraction remains larger than the volume fraction throughout the simulation. We find that the first multistreaming regions occur at later times in the WDM simulation, consistent with the slower collapse of larger scales. The truncated initial power spectrum leads to an overall lower fraction of multistreaming volume and mass at any time during the entire simulation. We notice that as expected, the first collapsed structures have linear-anisotropic velocity dispersion, before a significant fraction of mass and volume becomes planar and spherically isotropic. Comparing the high and low resolution results we notice that even for the 300WDM1 and 300WDM2 simulations, the mass fractions at high redshifts depend on the resolution. This discrepancy becomes naturally larger in the case of the 300CDM simulation, as new small-scale perturbations are added when the resolution of the simulation is increased. Since these small fluctuations are the first to collapse, the difference becomes particularly evident at early times.

\begin{figure*}
 \includegraphics[width=\linewidth]{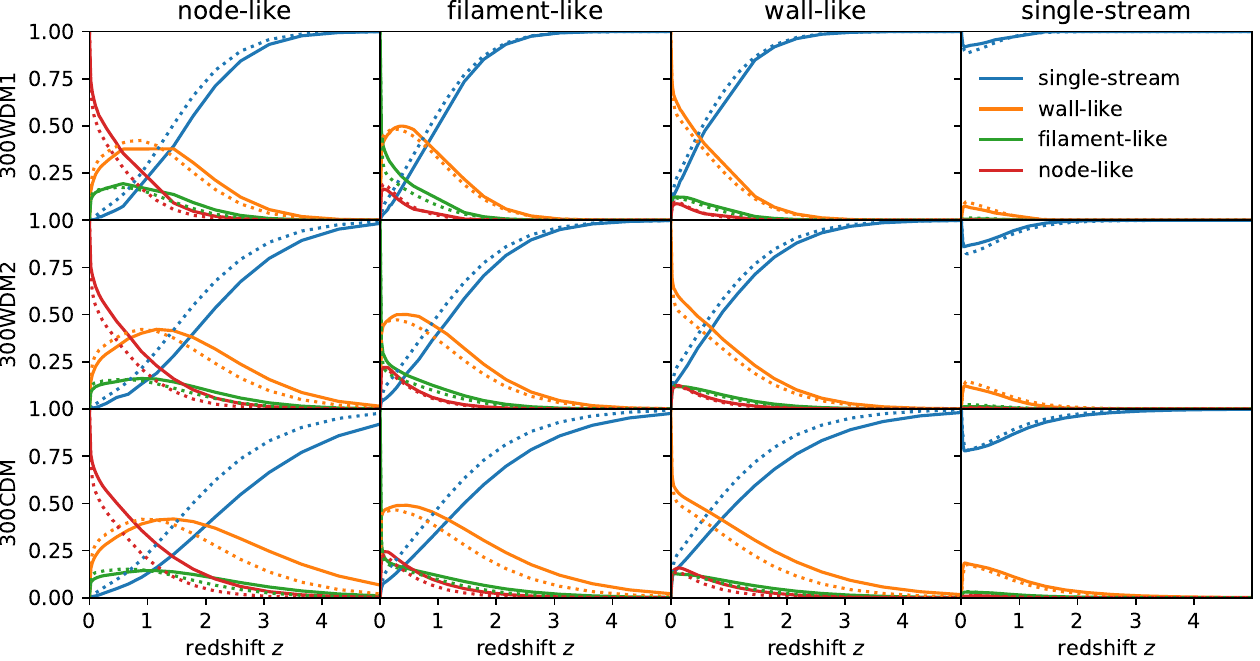}
 \caption{Environmental origin of the mass in node-like (first column), filament-like (second column), wall-like (third column) and single-stream (right column) environments at $z=0$ for the three different dark matter variants (rows). We show the mass fractions by environment at redshift $z$ that are in the target environment at redshift $z=0$. The results from the xDM\_10 simulations are plotted with solid and their lower resolution xDM\_9 counterparts with dotted lines.}
 \label{fig:mass_history}
\end{figure*}

In order to better understand the evolution of the environments, we follow the dark matter particles that reside in the various environments at $z=0$ back in time. At each earlier snapshot we compute the relative mass fractions of the environments. The results up to $z=5$ are shown in \autoref{fig:mass_history}.
Starting from the single-stream regions, we find that their progenitors were mostly single-stream regions. Wall-like regions mainly feed from single-stream regions, with a small fraction passing through cells with planar or spherical anisotropy. As predicted by the anisotropic collapse model, walls collapse to filaments, hence the progenitors of filament-like regions were mainly in wall-like regions. For the node-like regions however, we find that many of their constituent particles appear to collapse directly from wall-like regions. Given the relatively low time and space resolution we have around haloes, this aspect should be investigated more closely in future work.

Among the progenitors of each environment, we find a small fraction of particles that have changed environment in the opposite way than predicted by the theory of anisotropic collapse. This fraction is larger in the colder simulations, with up to 20\% of the void progenitors passing through multistreaming cells and 25\% of the filament progenitors coming from node-like regions. This is most likely due to the finite resolution of the rasterisation grid. If the cell encloses multiple environments (either due to the small size of the collapsed region in CDM, or at a boundary), a particle still in a lower level in the collapse hierarchy might be attributed to a higher level environment dominating the cell (e.g. a void particle assigned to a wall-cell). This issue could be avoided by evaluating the tessellation directly at the particle positions.

The scale of the WDM cut-off does not change the qualitative results of mass flowing from the single-stream regime to wall-, filament- and finally node-like structures. However, the fraction of mass flowing in the opposite direction increases with the amount of small-scale structure, further indicating that this is an effect of the rasterisation cell size. 
Changing the resolution of the simulation does not alter the mass fractions of the progenitors significantly. However, since the measured collapsed mass fraction at a given time is lower at lower resolutions if the perturbations are not fully captured (see discussion above), the mass transport from one environment to the next is also delayed. This is especially evident for the 300WDM2 and 300CDM simulations as well as the node-like environments in the 300WDM1 case. 

This so-called \emph{mass transport} across the cosmic web has been studied in detail e.g. by \citet{Cautun2014} using the \textsc{nexus+} algorithm. Qualitatively, these authors reported similar trends in mass flowing from voids to walls to filaments and finally to haloes, including significant reverse flows which they attributed to incorrect identification of environments in underdense regions. Their results differ in the timescale of matter transport through the environments, with collapsed environments being overall more ``stable'' between $z=2$ and today, whereas we find large mass fractions in uncollapsed regions. This is most likely tied to the unconverged collapsed mass fraction in cold simulations discussed above. Furthermore, they observe a filamentary mass fraction of $\sim$80\% as the progenitors of nodes, whereas in our measurements only $\sim$20\% of the node mass has previously been in filaments. This most likely is connected to our overall lower filamentary mass fractions and using a different segmentation of the anisotropy triangle will most likely lower the discrepancies.


\subsection{Alignment of the cosmic velocity dispersion $\sigma^2_{ij}$} \label{sec:alignment}
In the previous section, we have measured the anisotropy of the velocity dispersion field. The characteristic direction of the anisotropy, i.e. the main axis of the tensor field in wall-like structures and the minor axis in filament-like structures, contains additional information on the axes of collapse. In \autoref{fig:visu2}, we visualise the directions in walls (middle) and filaments (bottom) in the 18~$h^{-1}$Mpc cube using the direction-colour encoding shown in the coloured unit sphere. For wall-like structures the main axis of the velocity dispersion is perpendicular to the structure itself, whereas in filament-like structures the minor axis is parallel to the filament as expected from the collapse history of these regions. We can already see by eye that the characteristic directions are consistent over large distances across an entire segment of the cosmic web. This remains true for  large structures even in the CDM simulation where the visualisations become however somewhat cluttered by small-scale structures. This long range consistency of the velocity dispersion tensor could be used in principle to further dissect the volume by unambiguously identifying individual walls and filaments. In this subsection, we will measure the typical extent of the alignment using a marked correlation function and measure its deviations induced by small-scale perturbations.

Since filaments and walls are to first order not spatially curved, we expect the characteristic directions of the velocity dispersion field to be consistent over their typical sizes. To quantify this alignment as a function of distance, we use so-called marked correlation functions. They extend the classical correlation function framework \citep[cf.][]{Peebles1980} to study the spatial clustering of (usually scalar) properties of objects, so-called \emph{marks} \citep[][]{Stoyan1984, Stoyan1994}. These marked correlation functions have already been successfully applied to study the clustering of galaxy properties \citep[e.g.][]{Beisbart2000, Sheth2005, Skibba2006, Skibba2013}, the self-alignment and tidal field alignment of cosmic voids \citep[][]{Platen2008}, and have more recently been suggested as a tool to constrain modified gravity models \citep{White2016}. 

The commonly used marked correlation function \citep[e.g.][]{Sheth2005} is defined as the ratio of the weighted to the unweighted correlation function,
\begin{equation}
  \mathcal{M}(r) = \frac{1+W(r)}{1+\xi(r)} \approx \frac{WW}{DD},
\end{equation}
where $1+W(r) = \sum_{ij} m(\bmath{x}_i, \bmath{x}_j)/\bar{m}$ is the sum over all objects $i, j$ at separation $r$ weighted by the mark function $m(\bmath{x}_i, \bmath{x}_j)$, which for scalar marks usually is its product, i.e. $m(\bmath{x}_i, \bmath{x}_j) = m_i m_j$. Analogous to the unweighted correlation function, the approximation $WW/DD$ (the ratio of the weighted to the unweighted pair counts) can be used to efficiently estimate the marked correlation function. If the marks are uncorrelated, $\mathcal{M}(r) = 1$. Correlated and anticorrelated marks manifest themselves as larger and smaller values respectively.

For our measurement, the marks are the velocity dispersion tensors (or more precisely its eigenvectors) defined in every volume cell, which we have to reduce to a scalar quantity in the mark function \citep[cf.][]{Beisbart2000}. We use the angle $\theta$ between the major or minor eigenvectors $\bmath{e}$ of $\sigma_{ij}^2$ at the volume elements located at $\bmath{x}_1$ and $\bmath{x}_2$ and since the eigenvectors are invariant under sign inversion, we define $m(\bmath{x}_i, \bmath{x}_j) = (\bmath{e}(\bmath{x}_1) \cdot \bmath{e}(\bmath{x}_2))^2 = \cos(\theta)^2$. Together with our grid-based data, this definition allows us to efficiently compute $\mathcal{M}(r)$ in Fourier space using
\begin{align} \label{eq:marked_corrf}
  \mathcal{F}(WW)(\bmath{k}) &= \frac{1}{\bar{m}} \int \mathop{d^3 \bmath{r}} \exp(i\bmath{k}\bmath{r}) \int \mathop{d^3 \bmath{r}_1} \left(e_i(\bmath{r}+\bmath{r}_1) e^i (\bmath{r}_1) \right)^2 \\
  &= \frac{1}{\bar{m}} \mathcal{F}(e_i e_j)(\bmath{k}) \cdot \mathcal{F}(e^i e^j)(-\bmath{k}),
\end{align}
with summation over $i,j = 1,2,3$. 

\begin{figure}
  \includegraphics[width=\columnwidth]{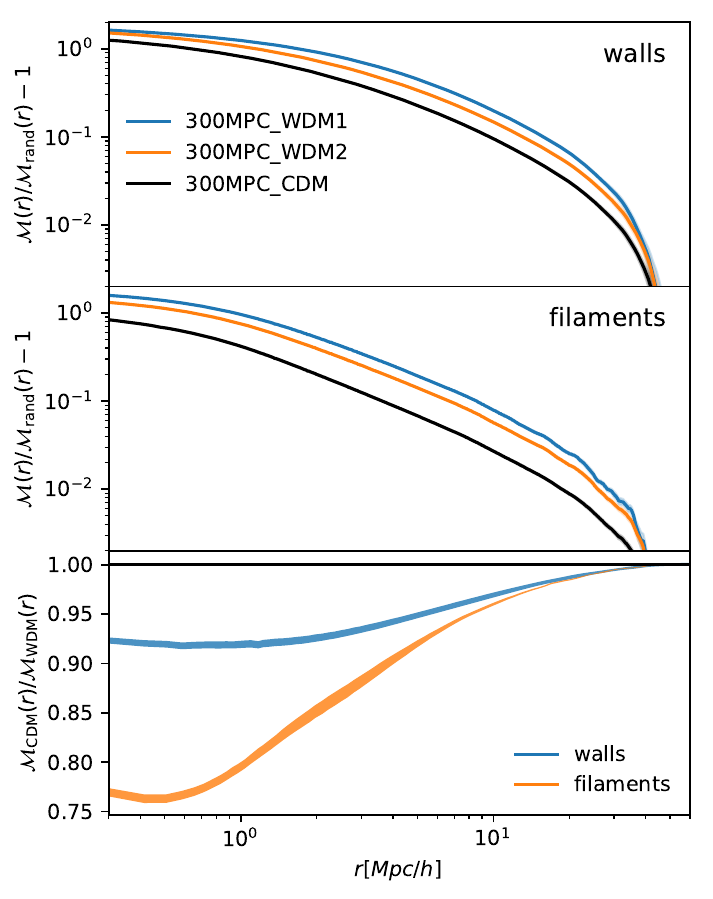}
  \caption{Relative strength of the marked correlation function (\autoref{eq:marked_corrf}) measuring the velocity dispersion alignment over the distance $r$ compared to a random field. We show the alignment of the major eigenvector field in wall-like structures (top) and minor eigenvector field in filament-like regions (middle) for the different simulations. Bottom: Relative strength of the marked correlation function in 300CDM\_10 compared to 300WDM1\_10, measured on the regions that are classified as walls and filaments in the 300WDM1\_10 simulation.}
  \label{fig:vd_dircorr}
\end{figure}

We are primarily interested in the excess of alignment compared to a random distribution of the eigenvectors. To measure this excess, we compute $\mathcal{M}_\mathrm{rand}(r)$ over five permutations of the velocity dispersion tensor field. Comparing $\mathcal{M}(r)$ with the variance of $\mathcal{M}_\mathrm{rand}(r)$ allows us to quantify the significance of our results. The results for the alignments of the major (minor) eigenvectors of $\sigma_{ij}^2$ in the wall-like (filament-like) collapsed regions are shown in \autoref{fig:vd_dircorr}. On small scales, the characteristic directions for both wall-like and filament-like structures are highly correlated, which is consistent with our expectation based on the spatial coherence of the tensors shown qualitatively in \autoref{fig:visu2}. The alignment decays with increasing distance and disappears for scales larger than $\sim 30$~$h^{-1}$Mpc which corresponds roughly to the typical extent of large walls and filaments found in simulations \citep{Cautun2014} and observations \citep[e.g.][]{Bond2010}. The alignment of the major eigenvector field in wall-like structures is generally stronger than the alignment of the minor eigenvector field in filament-like regions, indicating that the orientation of the velocity dispersion field is more consistent in the early stages of collapse (along walls) than in regions that have evolved further in the collapse hierarchy (along filaments). 

Going from the WDM to the CDM simulation, the alignment correlations remain highly significant but decrease at all scales as expected. Small-scale initial perturbations on the one hand create additional collapsed structures in previously uncollapsed regions, and on the other hand isotropise the smooth large-scale structures by causing them to fragment into smaller filaments and haloes. In order to measure this``isotropisation'' we compare the mark correlation function measured on the wall-like and filament-like support of the 300WDM1\_10 simulation between the CDM and WDM1 realisations. The result is shown in the lower panel of \autoref{fig:vd_dircorr}. On short separations the marked correlation function in the 300CDM simulation is ~8\% lower in wall-like regions and ~25\% lower in filament-like regions. As the marked correlation approaches the random distribution on larger scales, the difference between the simulations vanishes.

Another way of locally quantifying the influence of small-scale structures on the large-scale walls and filaments is to compare the strength of the velocity dispersion along and perpendicular to the structure and how it changes by adding perturbations. For this we define the angle between the two velocity dispersion components as $\tan(\alpha) = \sigma^2_\parallel / \sigma^2_\perp$, where the orientation is defined by the eigenvectors of $\sigma_{ij}^2$ in the WDM1 simulation. To compute the parallel and perpendicular components, we transform $\sigma_{ij, \mathrm{xDM}}^2$ to the eigenframe of $\sigma_{ij, \mathrm{WDM1}}^2$ and use its diagonal components. 

\begin{figure*}
  \includegraphics[width=\linewidth]{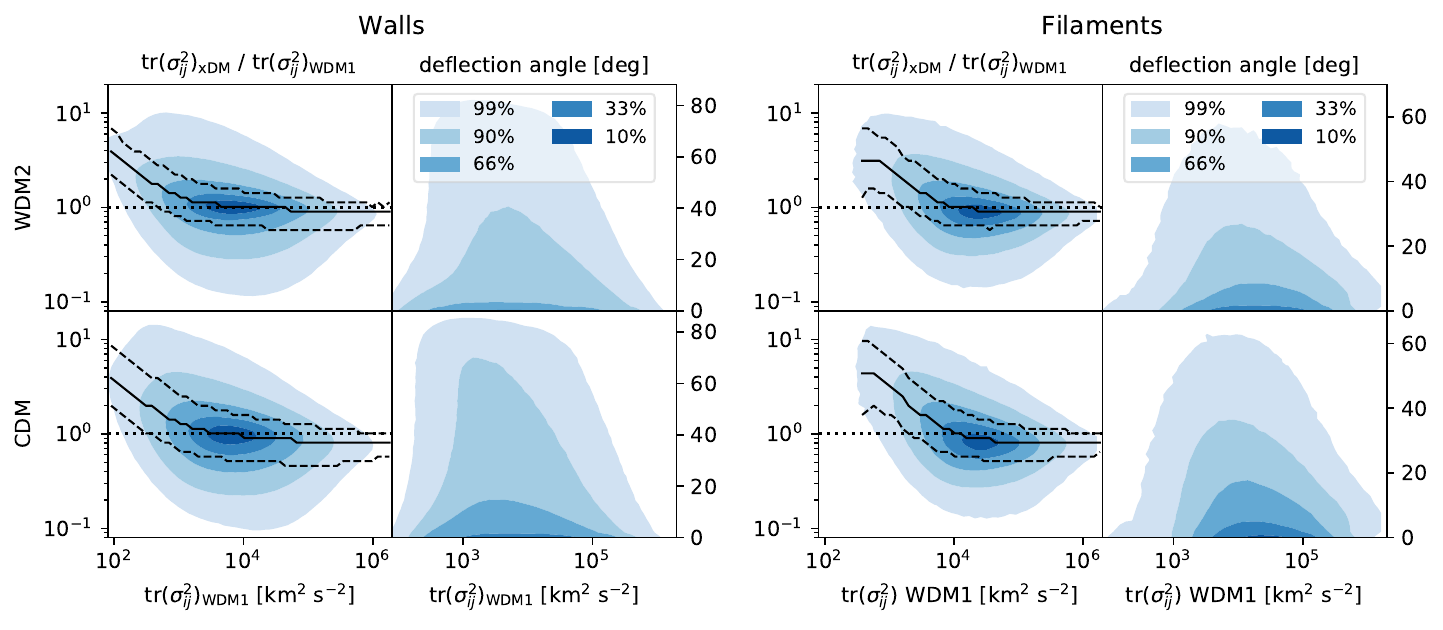}
  \caption{Comparison of the velocity dispersion in wall-like regions (left) and filament-like regions (right) between the 300WDM1 and 300WDM2 and 300CDM simulation respectively. We measure the relative change in amplitude of the velocity dispersion and the change of angular alignment of the velocity dispersion defined as the angle spanned by the velocity dispersion along and perpendicular to the wall (filament) $\tan(\alpha) = \sigma^2_{||} / \sigma^2_\perp$. The directionality of the structures is defined by their characteristic eigenvector of the velocity dispersion tensor in the 300WDM1\_10 simulation. The contour levels include 99\%, 90\%, 66\%, 33\% and 10\% of the volume elements respectively and the lines show the median and 66\% intervals at fixed velocity dispersion amplitude.}
  \label{fig:wall_fil_comp}
\end{figure*}

In \autoref{fig:wall_fil_comp} we compare the point-wise differences in amplitude and alignment angle between the 300WDM1 simulation and the 300WDM2 and 300CDM simulations respectively for wall-like and filament-like regions. Since the amplitude of the velocity dispersion is connected to the scale of the structure (cf. \autoref{sec:amplitude}), we plot the distributions as a function of $\mathrm{tr}(\sigma^2_{ij})$ to check for additional biases. We find that the velocity dispersion in regions with low amplitude in 300WDM1 is generally enhanced in both filaments and walls, but less affected by the small-scale structure if the velocity dispersion is already large. Overall, both the amplitude of $\sigma^2_{ij}$ and the alignment are highly consistent in most of the wall-like and filamentary volume.


\subsection{Correlation with the gravitational tidal field}
The Zel'dovich approximation (cf. \autoref{sec:theory}) predicts that anisotropic collapse is dictated by the large-scale tidal field tensor
\begin{equation}\label{eq:tf}
  T_{ij} = \frac{\partial^2\phi}{\partial x_i \partial x_j},
\end{equation}
where $\phi$ is the gravitational potential. Since the large-scale gravitational potential is constant at linear order, the large-scale tidal field remains correlated with the collapse evolution of the cosmic web. This fact has been used previously to classify the cosmic volume into the void-wall-filament-node morphology \citep{Hahn2007}. Since in the ZA the particle velocities before shell-crossing obey $\mathbf{v}\propto\boldsymbol{\nabla}\phi$ (this relation can be extended to the early non-linear regime, see e.g. \citet{Chodorowski2002, Ciecielg2003}, but the corrections remain small if the velocity field is filtered on sufficiently large-scales), a similar argument implies that before shell-crossing (or smoothed on large-scales), the velocity divergence tensor of the mean velocity field also reflects these cosmic web environments (used in the V-web classification of \citealt{Hoffman2012}). After shell-crossing, this inflow pattern gives rise to the anisotropic dispersion that we discuss and quantify in this paper.
 
We thus want to ask next whether we can recover this expected correlation between the two tensor fields: the tidal field as the dynamic origin of anisotropic collapse, and the large-scale velocity dispersion tensor as the result and signature of anisotropic collapse.

We compute the tidal field tensor of the large-scale structure from the smoothed density field measured by the CIC deposition of the dark matter particles. We use a Gaussian kernel with radius $r_s = 1$~$h^{-1}$Mpc to filter out small-scales. The qualitative results are not sensitive to $r_s$ and the measured alignment only drops significantly for $r_s < 500$~$h^{-1}$kpc and $r_s > 4$~$h^{-1}$Mpc.
 The tidal field can be conveniently derived in Fourier space as $\hat{T}_{ij}(\bmath{k}) = - 4 \upi G \, (k_i k_j) k^{-2} \, \hat{\rho}_s(\bmath{k})$. We compute the eigenvalue decomposition in each cell to obtain the principal axes of the tidal field and measure the angle $\theta$ between its eigenvectors and the eigenvectors of the velocity dispersion tensor.
\begin{figure}
  \includegraphics[width=\columnwidth]{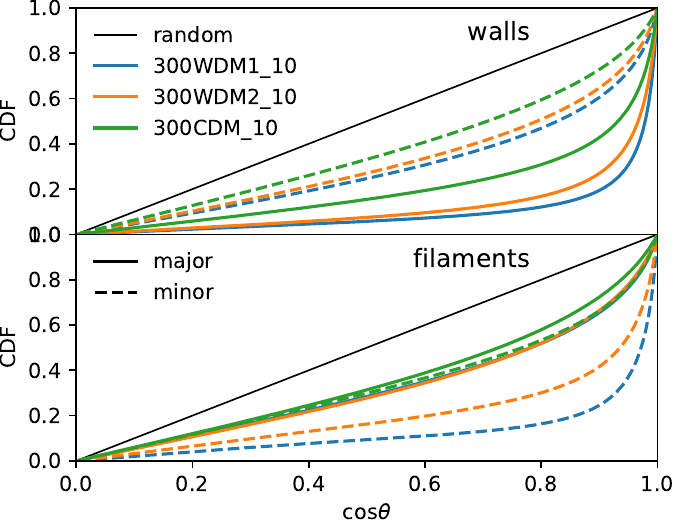}
  \caption{Cumulative distribution function of angles between the major (solid) and minor (dashed) eigenvectors of $\sigma_{ij}^2$ and the tidal field tensor. The tidal field has been computed from the smoothed density field ($l_S = 1$~$h^{-1}$Mpc) at redshift $z=0$. The distributions are plotted separately for the wall-like (top) and filament-like (bottom) regions.} 
  \label{fig:tf_align}
\end{figure}
\autoref{fig:tf_align} shows the cumulative distribution function (CDF) of the angles between the major-major and minor-minor eigenvectors of the two fields within the wall-like and filament-like regions. For a purely random distribution, $\cos \theta$ is uniformly distributed over $[0, 1]$. We notice a strong positive alignment for both vector fields in all simulations, with the major eigenvectors being stronger aligned in wall-like regions, and minor eigenvectors in filament-like regions. This follows directly from the distinct nature of the major and minor directions in walls and filaments as discussed previously. As expected, the two vector fields have the highest alignment in the 300WDM\_10 simulation with 73\% (67\%) of the major (minor) eigenvectors in wall-like (filament-like) regions deviating by less than 20 degrees from their counterpart. The deviations in the velocity dispersion alignment discussed in the previous section lower this alignment in the colder simulations, especially in filamentary regions that have progressed further in the collapse hierarchy (43\% of the major eigenvectors in walls and 33\% of the minor eigenvectors in filaments within 20 degrees in the 300CDM\_10 simulation). However, the alignments remain highly significant.


\section{Discussion and conclusions} \label{sec:conclusions}

During the anisotropic gravitational collapse of cosmic structure from cold initial conditions, kinetic energy is absorbed after shell-crossing into velocity dispersion, or stress, and provides the effective pressure that resists gravity in virial equilibrium. Due to the collisionless nature of dark matter, there is no microscopic process that renders this velocity dispersion isotropic so that it must retain some memory of anisotropic collapse in the cosmic web on large scales. On the other hand, the collapse of smaller structures always precedes the collapse of larger structures in the hierarchical structure formation scenario of CDM. One thus expects that smaller scale perturbations always act to increase the isotropy of the dispersion tensor. 
In order to disentangle this influence of small-scale structures from other non-linear, or even numerical effects, we always considered three simulations that all start from the same random phases. In addition to a vanilla CDM simulation, where the perturbation spectrum is effectively unresolved, we also include two simulations which start from initial conditions with suppressed small-scale fluctuations (exactly like in WDM scenarios). 

For these three flavours of simulations, we computed the velocity dispersion tensor directly from the CDM distribution function that we reconstructed using the phase-space sheet tessellation technique \citep[cf.][]{Abel2012, Shandarin2012}, which has previously been shown to yield highly accurate results for velocity fields \citep{Hahn2014}. We characterise the magnitude of the dispersion tensor through its trace value, and the anisotropic nature through three characteristic dimensionless numbers, the linear, planar and spherical anisotropy (\autoref{sec:anisotropy}). The relative dominance of one over the others of these numbers allows a parameter-free definition of wall-like, filament-like and node-like environments. This is a consequence of the collapse along one, two or three directions, causing large velocity dispersion along precisely those axes. Voids, in contrast, are characterised by the vanishing of the dispersion tensor since, in this picture, they are still simply single-stream regions. 

Our main results regarding the one and two-point statistics of the dispersion tensor can be summarised as follows:
\begin{enumerate}
  \item the amplitude of the velocity dispersion at $z=0$ is correlated with density in high density regions, and anticorrelated in collapsed regions below a simulation dependent threshold ($\delta < 4$ in the warmest simulation and $\delta < 1$ for the CDM simulation). For $\delta>0$ we find for the amplitude a scaling $\mathrm{tr}(\sigma_{ij}^2) \propto (1+\delta)^\alpha$ with $\alpha \sim 0.5 - 1$.
  \item the anisotropy of $\sigma_{ij}^2$ is strongly correlated with density, with environments below $\delta \sim 10$ having a strong linear anisotropy and turning isotropic at higher densities.
  \item the velocity dispersion power spectrum is proportional to the linear theory density power spectrum on large scales, but decays faster than the non-linear matter spectrum on small scales.
  \item the velocity dispersion -- density cross-spectrum behaves similarly on large scales but becomes negative above $k \gtrsim 3$~$h$Mpc$^{-1}$.   
  \item the velocity dispersion tensor field is spatially correlated not only in magnitude, but also in direction. This correlation extends over the typical size of filaments and walls in the cosmic web and is in agreement with the model of anisotropic collapse causing a consistent alignment of $\sigma_{ij}^2$ over a collapsed large scale mode.
  \item the velocity dispersion tensor  is very well aligned with the large-scale tidal field tensor, which is responsible for the anisotropic collapse on large scales. This implies that large-scale random motions in shell-crossed regions still reflect their origin from anisotropic collapse. 
\end{enumerate} 

A large amount of studies have been devoted to dissect the cosmic web into distinct components in $N$-body simulations \citep[e.g.][and many more]{Hahn2007, AragonCalvo2007, AragonCalvo2010, Sousbie2011, Hoffman2012, Falck2012} with a wide variance of results on both the volume and mass occupying the various structures \citep[see e.g.][for a comparison of the various methods]{Libeskind2018}. Typically, these methods require either the introduction of a filter scale (owing to the multi-scale nature of the cosmic web, in which small haloes sit inside filaments that sit inside large-scale walls) or some tuning of multi-scale filter parameters. The velocity dispersion tensor allows a parameter free determination of the same environments and is directly motivated by the anisotropy of large-scale gravitational collapse. We can directly confirm previous results that
\begin{enumerate}
  \item mass predominantly flows from voids to walls to filaments and finally to haloes, in agreement with expectations from anisotropic collapse,
  \item nodes occupy the densest regions, followed by filaments and walls. The measured mean densities are however highly dependent on the amount of small-scale structure that can be captured by the resolution of the simulation, and generally decrease with increased resolution.
\end{enumerate} 
We expect that these results can give important insights into the anisotropic nature of gravitational collapse and the emergence of anisotropic stress in the cosmic web which are of great importance in effective perturbative models of large-scale structure formation and evolution, but also in the modelling of redshift space distortions in cosmological observations. A further interesting future application is to investigate the statistics of the ``coldness'' of local Hubble flows \cite[e.g.][]{Karachentsev2002,AragonCalvo2011}.

\section*{Acknowledgments}
We thank Cornelius Rampf for comments and suggestions and Olivier Durif for first preliminary investigations regarding the anisotropy of the cosmic velocity dispersion in his master thesis at OCA.
We thank the anonymous referee for valuable suggestions that helped to improve the presentation of this article.
Simulations were carried out using \textsc{gadget-2} \citep{Springel2005}. Ternary diagrams were created using \texttt{python-ternary} \citep{PythonTernary}.
MB and OH acknowledge funding from the European Research Council (ERC) under the European Union's Horizon 2020 research and innovation programme (grant agreement No. 679145, project 'COSMO-SIMS').




\bibliographystyle{mnras}
\bibliography{bibliography}



\appendix

\section{Analytical model for one dimensional plane wave collapse} \label{sec:1dtoy}
In this section we construct a very rough model to estimate the velocity dispersion of a plane wave right after the collapse time $a_\times$. A more thorough treatment including post-collapse corrections can be found in \citet{Taruya2017}.
In order to be able to invert $ \bmath{x} = \bmath{x}(\bmath{q})$ analytically after shell-crossing,  we expand the plane wave perturbation with mode $k$ around $q=0$ to the lowest order that leads to collapse, 
  \begin{equation}
    x(q, a) = \left[1-\frac{D_+(a)}{D_+(a_\times)}\right]q + \frac{k^2}{6} \frac{D_+(a) }{ D_+(a_\times)}q^3+\mathcal{O}(q^5).
  \end{equation}
  This expression has one real root for $a<a_\times$ and three for $a>a_\times$, corresponding to the dark matter sheets crossing $x=0$. In catastrophe theory (e.g. \citealt{Poston1978}, but see also \citealt{Arnold1982,Hidding2014}), this is also called a \emph{normal form}, describing the topological structure of the first shell-crossing, and such a system is referred to as the \emph{cusp catastrophe}. We include this Taylor expansion in \autoref{fig:1dcollapse}. It tightly follows the ZA around $q=0$ but starts to deviate further away from the centre of collapse. This causes the approximation to underestimate the velocity dispersion (compared to Zel'dovich) at late times.
    
  Focusing on the centre $x=0$ of the perturbation, we can express the velocity dispersion (in comoving velocity units) as
  \begin{equation} \label{eq:1d_vd_approx}
    \sigma_\mathrm{c}^2(x=0, a) = \frac{3}{k^2}\left(1 - \frac{D_+(a_\times)}{D_+(a)}\right) \left(\frac{\dot{D}_+(a)}{D_+(a)}\right)^2,
  \end{equation}
  for $a >= a_\times$. 
  To get an estimate on $\sigma^2_c$ immediately after collapse, we evaluate this equation at $a = a_\times(1+\Delta a)$ with $\Delta a \ll 1$. Furthermore, we assume an Einstein de-Sitter (EdS) universe ($\Omega_m = \Omega_\mathrm{tot} = 1$), for which the growth factor scales as $D_+(a) = a$ and $\dot{D}_+(a) = \dot{a} = H_0^2 a^{-1/2}$ and obtain
  \begin{equation}
    \sigma_c^2 \left(x=0, a_\times(1+\Delta a)\right) = \frac{3 H_0^2}{k^2}  \frac{\Delta a}{1 + \Delta a} \left( a_\times (1+\Delta a) \right)^{-3}.
  \end{equation}
  Recalling the shell-crossing time of a plane wave $a_\times = A^{-1} k^{-2}$, we find that at fixed $\Delta a$, the comoving velocity dispersion $\sigma_c^2 \propto A^3 k^4$. The typical amplitude of the potential is dependent on the scale $k$ and related to the matter power spectrum as $A(k) \sim (P_{\delta\delta}(k) k^{-4})^{1/2}$. We therefore expect $\sigma_c^2 \propto P_{\delta\delta}^{3/2} k^{-2}$, which implies that for scales sufficiently smaller than the Hubble horizon at radiation-matter equality, $k > k_\mathrm{eq}$, small-scale perturbations are expected to have lower velocity dispersion at a fixed time after shell-crossing. Of course, this is only a very rough model of the actual physics, neglecting the three dimensional nature of collapse, the presence of perturbations on all scales and post-collapse corrections.

\section{Comparison of cosmic web environments detected by different finders} \label{sec:cosmicweb_comparison}
To compare our method with other cosmic web finders, we  briefly discuss the density dependence of the environments detected by different cosmic web finders. We use the public data from the cosmic web comparison paper by \citet{Libeskind2018}, which includes a $N$-body simulation\footnote{Simulation set-up: 200 $h^{-1}$Mpc box, $512^3$ particles with $\Lambda$CDM cosmology and \citet{Planck2015} parameters} snapshot at $z=0$ and the classification of the cosmic web environments on a $100^3$ grid. The public data includes various classification techniques and we refer the reader to \citet{Libeskind2018} for further information and references.

\begin{figure*}
  \includegraphics[width=\linewidth]{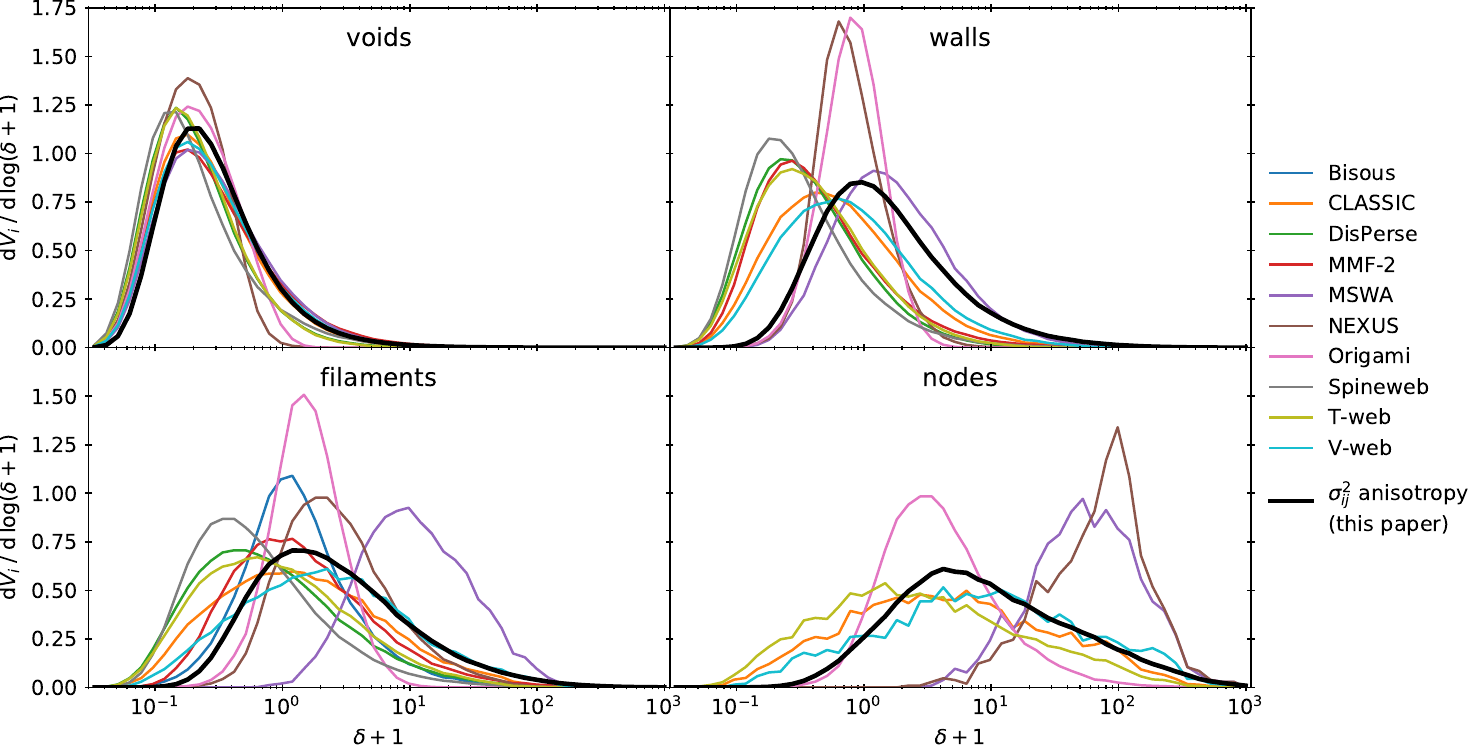}
  \caption{Comparison of the density contrast $1+\delta$ distribution (normalized) of the cosmic web environments found by different cosmic web finders (note that not all environments are classified by every finder). The comparison data has been obtained from the data published together with the cosmic web comparison paper by \citet{Libeskind2018} containing more details about the individual classifiers.}
  \label{fig:webfinder_comparison}
\end{figure*}

We compute the volume averaged density field using the \textsc{dtfe} code \citep{Schaap2000, vandeweygaert2009, Cautun2011} on the same $100^3$ grid for the provided snapshot as well as on a $256^3$ grid for the CDM300\_512 simulation at $z=0$. \autoref{fig:webfinder_comparison} shows the measured density distribution of each environment. Note that not all classifiers detect every environment and hence some lines are missing from some of the panels. Additionally, we compute the volume and mass fractions of each environment and its mean density. The values are listed in \autoref{tab:volmass_comparison}.

Overall, the cosmic web environments detected via the velocity dispersion anisotropy discussed in this paper are consistent with the range of density distributions and mass and volume fractions from existing methods. As already reported in \citet{Libeskind2018}, the measured quantities of the cosmic web regions highly depend on the applied definition. The mass and volume fractions measured in this paper agree best with the \emph{MultiStream Web Analysis} \citep[MSWA][]{Shandarin2012, Ramachandra2015} method as we have already noted in \autoref{sec:anisotropy}. However, filaments and nodes extend to lower densities than the ones identified with MSWA and their density distributions are more similar to the environments identified with V-web \citep{Hoffman2012}, MMF-2 \citep[filaments only, ][]{AragonCalvo2014}, CLASSIC (see \citet{Libeskind2018} for method and further references) and T-web \citep{Hahn2007, Gottlober2009}.

\begin{table*}
  {\scriptsize
  \begin{tabular}{@{}l|cccc|cccc|cccc@{}}
    Method & \multicolumn{4}{c|}{volume fraction} & \multicolumn{4}{c|}{mass fraction} & \multicolumn{4}{c}{mean density} \\
    \midrule
    & uncollapsed & walls & filaments &  nodes & uncollapsed & walls & filaments &  nodes & uncollapsed   & walls         & filaments     & nodes\\
    & \%          & \%    & \%        & \%     & \%          & \%    & \%        & \%     & $\bar{\delta}$& $\bar{\delta}$& $\bar{\delta}$& $\bar{\delta}$\\
    \midrule
    Bisous     &  --  &  --  & 12.1 &  --  &  --  &  --  & 31.0 &  --   &  --  &  --  &  2.6 &  --  \\
    CLASSIC    & 70.3 & 23.8 &  5.3 &  0.6 & 31.1 & 32.2 & 23.4 & 13.3  &  0.4 &  1.4 &  4.4 & 21.1 \\
    \textsc{DisPerse}   & 38.8 & 37.3 & 23.9 &  --  & 12.5 & 25.2 & 62.3 &  --   &  0.3 &  0.7 &  2.6 &  --  \\
    MMF-2      & 73.3 & 19.0 &  7.8 &  --  & 47.9 & 19.8 & 32.3 &  --   &  0.7 &  1.0 &  4.2 &  --  \\
    MSWA       & 90.3 &  8.8 &  0.7 &  0.1 & 49.5 & 27.9 & 13.0 &  9.6  &  0.5 &  3.1 & 17.7 & 84.5 \\
    \textsc{nexus}      & 65.7 & 22.8 & 11.3 &  0.1 & 14.7 & 21.5 & 52.8 & 11.0  &  0.2 &  0.9 &  4.7 & 95.0 \\
    \textsc{origami}    & 73.8 & 12.3 &  6.4 &  7.5 & 19.5 & 11.5 & 11.9 & 57.1  &  0.3 &  0.9 &  1.9 &  7.6 \\
    Spineweb   & 33.2 & 30.7 & 36.1 &  --  & 14.7 & 22.5 & 62.8 &  --   &  0.4 &  0.7 &  1.7 &  --  \\
    \textsc{T-web}      & 42.5 & 41.3 & 14.9 &  1.3 & 13.3 & 31.3 & 37.6 & 17.8  &  0.3 &  0.8 &  2.5 & 14.0 \\
    \textsc{V-web}      & 78.7 & 18.1 &  3.0 &  0.2 & 39.2 & 32.5 & 20.3 &  8.1  &  0.5 &  1.8 &  6.9 & 35.4 \\
    $\sigma_{ij}^2$ anisotropy & 91.3 & 7.4 & 1.0 &  0.3 & 51.1 & 30.4 & 8.2 & 10.3 & 0.6 &  4.1 & 8.0 & 34.5\\
  \end{tabular}
  \caption{Mass and volume fractions and the mean density of the cosmic web environments. The comparison data has been obtained from the data published together with the cosmic web comparison paper by \citet{Libeskind2018}. The data from the velocity dispersion anisotropy classification described in this paper has been computed from the 300CDM\_512 simulation. Since we're using the \textsc{dtfe} density estimation for comparability, the mass fractions and mean densities reported for the 300CDM\_512 simulation show a small deviation from the data in \autoref{tab:volmass_dist}.}
  \label{tab:volmass_comparison}
  }
\end{table*}

\bsp	
\label{lastpage}
\end{document}